\documentclass[journal,10pt,a4paper]{IEEEtran}
\usepackage{mathtools}
\usepackage{verbatim}   
\usepackage{graphicx}
\usepackage{listings}
\usepackage{color}
\usepackage{comment}
\usepackage{xcolor}
\usepackage{lipsum} 
\usepackage{breqn}
\usepackage{epsfig}
\usepackage{epstopdf}
\usepackage{subfigure}
\usepackage{placeins}
\usepackage{url}
\usepackage{colortbl}
\usepackage{makecell}
\usepackage{rotating}
\usepackage{multirow}
\usepackage{soul}
\usepackage{cite}
\usepackage{hhline}
\usepackage{times}
\usepackage{fancyhdr,graphicx,amsmath,amssymb}
\usepackage[ruled,vlined]{algorithm2e}

\begin{document}
	\pagestyle{empty}

	\title{Transfer learning-aided, Artificial Intelligence Approach to Evaluate Optical Properties of Materials for Classical and Quantum Photonic Applications}
	
	\author{Mohammad Reza Habibi $^{1,*}$, \textit{} Luca Sapienza $^{2,*}$, and Juan C.~Vasquez $^{3}$

1- AAU Build, Aalborg University, A.C. Meyers Vænge 15, 2450 København SV, Denmark. e-mail: mre@build.aau.dk

2- Department of Engineering, University of Cambridge, Cambridge, CB3 0FA United Kingdom. e-mail: ls2052@cam.ac.uk

3- AAU Energy, Aalborg University, Pontoppidanstræde 111, 1025, 9220 Aalborg Øst, Denmark. e-mail: juq@energy.aau.dk

    ({*}- Corresponding authors: Mohammad Reza Habibi and Luca Sapienza)

  

	}
\markboth{}%
{Shell \MakeLowercase{\textit{et al.}}: Bare Demo of IEEEtran.cls for IEEE Journals}

	\maketitle

\begin{abstract}
In this study, an artificial intelligence-assisted evaluation of the refractive indices of different materials is performed. The evaluation of the refractive index is necessary to design photonic devices for various classical and quantum photonic applications, and to theoretically analyse experimental results. The refractive index is one of the crucial parameters that affect the interaction of light with matter. Although the refractive index can be measured experimentally, its estimation via computational means provide the advantage of not requiring costly instrumentation, trained researchers and could avoid potential laboratory-based uncertainties. Therefore, in this study, artificial intelligence is implemented to estimate the optical constants of two types of materials, i.e., noble metals, and semiconductors. To this end, an artificial neural network is used to estimate the real and the imaginary parts of the refractive index of a specific material [i.e., copper (Cu) for noble metals, and silicon (Si) for semiconductors] for various wavelengths of light. Then, the knowledge of the trained neural network is employed to initiate and train new neural networks to estimate the refractive indices of other materials [i.e., gold (Au) for noble metals, and germanium (Ge), gallium arsenide (GaAs), and indium phosphide (InP) for semiconductors]. For the training of a neural network, various methods can be deployed. However, the application of the training can be a time-consuming process. A transfer learning approach is here implemented to provide a faster training solution for the problem under study. The implementation of transfer learning allows training the neural networks for a new material with higher speed or accuracy when compared to non-knowledge transfer-based training. Our results show the efficiency of the proposed technique in accurately evaluating the real and imaginary parts of the refractive index of different materials. Based on the obtained results, the achieved regression coefficient for estimation of the real part and the imaginary part of Au are 0.997 and 1, respectively. Furthermore, the obtained regression coefficients related to Ge, GaAs and InP are more than 0.998 and 0.992, for the real part and the imaginary part of the refractive, respectively.
\end{abstract}

	\section{Introduction}

\IEEEPARstart{Q}{\lowercase{uantum}} computing can be studied and implemented in different applications, e.g., drug discovery \cite{9520764, 10082989, PYRKOV2023103675, 10466774, electronics12112402, bikku2024improved}, power systems \cite{habibi2025electrical, 9831167, 9423668, electronics11182919, 9737359, 10144277, GOLESTAN2023584}, etc. Quantum computing schemes based on quantum light  have been proposed \cite{knill2001scheme, franson1998optical}, and the interaction of light with different materials is at the basis of many quantum technology applications. The refractive index is a complex index including a real and an imaginary part that determines the behaviour of light when interacting with materials. The refractive index \cite{ShyamSingh_2002} plays an important role in optics, as well as in the design of both classical and quantum systems\cite{ANANI2008570} such as LED and thermal sources, and single or entangled photon sources, respectively.

For example, to design AlGaN-based LEDs, the refractive index engineering can be considered as a parameter, which can lead to improve the performance of the system \cite{zheng2024refractive}. Furthermore, the design of an area blackbody can be affected by the refractive index of the selected material, e.g., in a multilayer anti-reflective coating-based area blackbody, a gentle decrement of the value of the refractive index from the top layer to the bottom layer can result in an enhancement of the emissivity \cite{ZHANG2022104311}. As another example, knowledge of the value of the refractive index is required for the operation of fibers, e.g., the difference of the refractive index between the cladding and the core is needed for high power fibers with single-mode operation \cite{reisner2022quantum}.  The relationship between the decoherence caused in atmospheric turbulence and parameters like refractive index constant, propagation distance, beam radius, and  wavelength for the case of orbital angular momentum entangled bi-photons has been investigated in \cite{ibrahim2014parameter}. Besides, an accurate control of the waveguide parameters (e.g., refractive index) is needed for the generation of polarization-entangled photons \cite{svozilik2012bragg} \cite{guerreiro2011production}, and the emission spectrum can influenced by parameters such as variations in refractive index induced by optical perturbations \cite{guerreiro2011production}. A refractive index perturbations-based mechanism has been proposed for controlled generation of entangled photons by the implementation of modulations (accelerating or superluminal) of refractive index in \cite{sloan2022controlling}. Also, for the case of marine 
environment, random variation of refractive index leads to a modulation of the light transmission \cite{9964005}.

The value of the refractive index can be measured experimentally, but, the measurement requires specific equipment like ellipsometry \cite{fujiwara2018spectroscopic}, that make the measurement process costly. In addition, experimental data can be affected by uncertainties, in particular when dealing with thin films and/or complex media. To overcome these issues, we present a technique based on computational methods and refractive index estimation supported by artificial intelligence (AI).

AI has been implemented in a wide range of applications e.g., cybersecurity of cyber-physical power electronics \cite{habibi2024artificial}, communication \cite{10183787, 10218388, 9039619}, cyber-attack detection and mitigation in power applications such as cyber-physical DC microgrids \cite{9591496, 9319658, 9244459}, image recognition \cite{9691259, 9129654, 9062591}, weather forecasting \cite{8864021, 8529196, 6730905, 8781906, 10285372, 9849651}, sonsorless strategies in power electronic-based power systems like microgrids \cite{9354239, 9291450}, prediction of electrical loads \cite{10111057, 10225522, electronics12071685}, stabilization of power applications such as DC microgrids \cite{9563239}, etc. In addition, a review on AI in quantum photonics and classical photonics (such as quantum optics, spectroscopy and chemometrics, as well as, computational imaging)  is presented in \cite{vernuccio2022artificial}. For more examples, an AI-based strategy-assisted (i.e., machine learning) quantum photonics is evaluated and discussed in \cite{kudyshev2020machine}, and a review on the implementation of AI in photonics (for example modeling and simulation) is reported in \cite{alagappan2022leveraging}. Furthermore, artificial neural networks as a model of AI have been deployed in quantum nanophotonics for the estimation of the local density of states in \cite{10088983}.

In this study, AI is implemented to accurately estimate the refractive index of different materials. One of the well-known AI-based methods is artificial neural network (ANN), which is a powerful application, that allows the estimation of the system's output using inputs without relying on any information about the entire system. In other words, to estimate a desired value in a system, an ANN requires only a dataset that includes inputs and the output of the system at some points. Then, without other extra information (e.g., mathematical-based information, knowledge about the entire losses, etc), an ANN can be trained to calculate the output of the system. Considering the advantage of the ANN (i.e., non-dependency to the mathematical and physical information of the studied system), this study implements ANNs to estimate optical constants of materials in different ranges of the wavelength of light, interacting with different media.

An ANN needs to be trained by a dataset with a proper size to tune the parameters of the network. Therefore, the first step before the implementation of an ANN is to provide a dataset including the inputs and the output of the ANN. After the preparation of the dataset, the main task is to train the ANN to make it ready for implementation in the system by tuning the parameters. Typically, the training process is an optimization process that tunes the parameters of the neural network by finding the optimal values of them using an optimization method. Generally, optimization strategies are time-consuming. Therefore, one of the main challenges regarding the training of an ANN is the long time needed for the optimization to achieve a good level of accuracy. The time is prolonged if the number of the iterations or the epochs increases. The required time depends to the complexity of the application and can be different, e.g., seconds or even shorter for very simple curve fittings like low-degree polynomial curve estimations, to hours or even days for complex models related to image processing. For an example, in a same application, the training process needs a longer time to train the model if the number of epochs changed from $10^5$ to $10^6$. To overcome this challenge, this paper uses transfer learning to reduce the number of iterations to achieve a better level of accuracy and to train a more accurate ANN model, therefore, achieving a more accurate model with the same number of iterations. For example, as it will be shown later, the validation performance (i.e., mean squared error) at epoch 100000 for the estimation of the real part of Au is about 0.0022 and 0.0015, without and with transfer learning, respectively. As another example, in this study, for the case of the real part of the refractive index of Au, the validation performance around 0.0022 occurs under the transfer learning at epoch number 29719, while the validation performance around 0.0022 occurs at epoch number 100000 without the implementation of transfer learning. Therefore, if the training time is assumed to be proportional to the number of epochs, the transfer learning reaches the validation performance 0.0022 about 3.36 times faster than without transferring the knowledge.

This technique allows to use and transfer the knowledge from an ANN that is trained for a specific material to another ANN that will be trained and implemented for another material. It is important to note that, the transfer learning will be tested under two conditions, i.e., a dataset with a small size including fewer samples (for semiconductors, and to evaluate whether good accuracy can be achieved even with a small dataset), and a larger dataset with more samples (for noble metals, and to assess its capability to process a larger dataset). In addition, a study is deployed to evaluate the noise reduction of the implemented model compared to a noisy dataset. As discussed in more detail below, based on the implemented study, the noise can be reduced using the deployed AI-based technique for both real and imaginary parts of the refractive index. In this paper, the following features are considered:

\subsubsection{Material data generation} A generation strategy will be done on a dataset with limited number of samples. The data generation strategy increases the size of the dataset using a linear interpolation-based data technique.

\subsubsection{Noise implementation} A uniform random number for each dataset is considered, which can play the role of measurement noises or errors to have more realistic results and to test the performance of transfer learning under a worse situation. Also, the AI-based strategy is evaluated using an index related to the existence of the noise, i.e., the ratio of signal power to  noise power, to evaluate the effectiveness of the proposed strategy to reduce the noise in the system.

\subsubsection{Knowledge transmission using transfer learning} To reduce the time duration that is needed for the training to achieve a more accurate model, the knowledge of the trained ANN for a specific material is transferred to a new ANN that will be trained for another material. This knowledge includes the architecture of the ANN (i.e., the number of inputs, the number of hidden layers, and the number of neurons of each hidden layer), and the tuned parameters of the ANN (i.e., the weighting coefficients between the neurons, and the bias factor of each neuron related to the trained ANN for one material). This feature allows us to reduce the time duration of the training related to the new ANNs for another material with more accuracy, i.e., less error in the calculations with a faster training process.

\subsubsection{Testing the new ANNs} The proposed strategy will be tested on datasets related to different materials, i.e., noble metals, and semiconductors. 

The refractive index is an important parameter to control the interaction of light with matter, and, consequently, the use of an incorrect or non-accurate value of that could lead to a decrease of the efficiency or performance in photonic-based devices. A data-driven method could be computationally expensive and can also rely on the preparation of a large dataset, that for the case of the refractive index could be time-consuming, complex, and costly. This study investigates the potential of transfer learning along with AI to evaluate the optical constants of materials.

Here, MATLAB is implemented to perform and investigate this study. Section II will describe the proposed transfer learning-based approach in more details covering the architecture of the neural networks, the knowledge preparation, and transferring of knowledge. Section III will present the results of the proposed strategy for noble metals. In this section, ANNs will be trained to calculate the real part and the imaginary part of Cu, and the knowledge of the trained ANNs will be used to provide neural networks for Au. It is important to note that, in Section III, the data generation strategy will be implemented to increase the size of the dataset. In addition, Section IV will examine transfer learning on semiconductor materials. Firstly ANNs will be trained for Si. Then, the knowledge of the trained ANNs of Si will be deployed to prepare ANNs for Ge, GaAs, and InP. In Section IV, a small dataset will be implemented with the same size as the original dataset to investigate the transfer learning-based strategy for materials for which only a limited data set is available, in order to further demonstrate the potential of the transfer learning-based technique to predict refractive indices of less well-known materials. Also, a technical comparison along with some results will be presented in Section V to highlight the advantages of transfer learning. In Section VI, the noise is investigated to evaluate the reduction of noise on the dataset. In Section VII, an ablation study will be conducted to evaluate the influence of transfer learning, and knowledge transfer from similar physically family material. 
A general discussion and our conclusions will be summarised in Section VIII and Section X, respectively. In addition, in Section X, some future directions related to this study will be suggested.

\section{Transfer Learning-based Strategy}
\label{Sec:Method}

The main goal of this study is to train ANNs for different materials to estimate the optical constants of them by reducing the required duration time of the training process or increasing the accuracy of the estimation. The optical properties of a material (i.e., refractive index) can be represented by a complex number as follows:

  \begin{equation}
N=n+ik.
\label{v11}
 \end{equation}

Two ANNs will be used namely ANN\textsubscript{r,x}, and ANN\textsubscript{i,x} for the estimation of the real part and the imaginary part of the refractive index for material $x$, respectively. In the next Subsection, the architecture of the ANNs will be discussed in more detail.

\subsection{Architecture of ANN\textsubscript{r,x}, and ANN\textsubscript{i,x}}

The first step is to have access to well-trained neural networks for a material. To this end, we chose copper and silicon as known noble metal and semiconductor, respectively. Here, we define ANN\textsubscript{r,Cu} and ANN\textsubscript{i,Cu} as the neural networks belong to the real part and the imaginary part of the refractive index for copper. Additionally, ANN\textsubscript{r,Si} and ANN\textsubscript{i,Si} are defined as the models corresponded to the real and the imaginary parts of the refractive index of silicon, respectively. In this work, for simplicity, the number of the hidden layers to connect the input layer to the output layer, for propagating data from the input layer to the output layer, is one. As will be shown later, with just one hidden layer, the ANNs have a high level of accuracy to estimate the desired data. In addition, the number of the inputs of the ANN is one, i.e., the input is only the photon energy. The parameters of ANN\textsubscript{r,x} and ANN\textsubscript{i,x} can be categorized as weighting co-coefficients and bias factors. If the number of the neuron of the hidden layer is $\alpha$, the weighting co-coefficients of the ANNs are as follows:

  \begin{equation}
W_{r,x,in}=\begin{bmatrix}
w_{r,x,in,1} & w_{r,x,in,2} & \cdots  & w_{r,x,in,\alpha}
\end{bmatrix},
\label{v12}
 \end{equation}

  \begin{equation}
W_{r,x,o}=\begin{bmatrix}
w_{r,x,o,1} & w_{r,x,o,2} & \cdots  & w_{r,x,o,\alpha}
\end{bmatrix},
\label{v13}
 \end{equation}

  \begin{equation}
W_{i,x,in}=\begin{bmatrix}
w_{i,x,in,1} & w_{i,x,in,2} & \cdots  & w_{i,x,in,\alpha}
\end{bmatrix},
\label{v14}
 \end{equation}

   \begin{equation}
W_{i,x,o}=\begin{bmatrix}
w_{i,x,o,1} & w_{i,x,o,2} & \cdots  & w_{i,x,o,\alpha}
\end{bmatrix},
\label{v15}
 \end{equation}

Where, $w_{r,x,in,j}$ and $w_{r,x,o,j}$ are the weighting coefficients between the input neuron and the $j^{th}$ neuron of the hidden layer as well as the $j^{th}$ neuron of the hidden layer and the output neuron of ANN\textsubscript{r,x}, respectively. In addition, $w_{i,x,in,j}$ and $w_{i,x,o,j}$ are the weighting coefficients between the input neuron and the $j^{th}$ neuron of the hidden layer of ANN\textsubscript{i,x}, respectively. The bias factors of ANN\textsubscript{r,x} and ANN\textsubscript{i,x} are defined as follows:

   \begin{equation}
B_{r,x,h}=\begin{bmatrix}
b_{r,x,h,1} & b_{r,x,h,2} & \cdots  & b_{r,x,h,\alpha}
\end{bmatrix},
\label{v16}
 \end{equation}

    \begin{equation}
B_{i,x,h}=\begin{bmatrix}
b_{i,x,h,1} & b_{i,x,h,2} & \cdots  & b_{i,x,h,\alpha}
\end{bmatrix},
\label{v17}
 \end{equation}

    \begin{equation}
B_{r,x,o}=\begin{bmatrix}
b_{r,x,o}
\end{bmatrix},
\label{v18}
 \end{equation}

 and, 

    \begin{equation}
B_{i,x,o}=\begin{bmatrix}
b_{i,x,o}
\end{bmatrix}.
\label{v19}
 \end{equation}

In \eqref{v16} and \eqref{v17}, $b_{r,x,h,j}$ and $b_{i,x,h,j}$ are the bias factor of the $j^{th}$ neuron of the hidden layer of ANN\textsubscript{r,x} and ANN\textsubscript{i,x}. In addition, in \eqref{v18} and \eqref{v19}, $b_{r,x,o}$ and $b_{i,x,o}$ are the bias factor of the output neuron of ANN\textsubscript{r,x} and ANN\textsubscript{i,x}, respectively. 

The input of the ANNs for metal $x$ is photon energy (i.e., $E_{\omega,x}$), and the output of the ANNs are $\overline{n}_{\omega,x}$ and $\overline{k}_{\omega,x}$ (i.e., the estimated values of the real part and the complex part of the optical property, respectively) for ANN\textsubscript{r,x} and ANN\textsubscript{i,x}. Based on the parameters of the ANNs, $\overline{n}_{\omega,x}$ can be calculated as follows:

    \begin{equation}
\overline{n}_{\omega,x}=f_{r,2}(\sum_{j=1}^{\alpha}(f_{r,1}(E_{\omega,x}\times w_{r,x,in,j}+b_{r,x,h,j})\times w_{r,x,o,j})+b_{r,x,o}).
\label{nomega}
 \end{equation}

 In addition, $\overline{k}_{\omega,x}$ can be calculated as follows:

     \begin{equation}
\overline{k}_{\omega,x}=f_{i,2}(\sum_{j=1}^{\alpha}(f_{i,1}(E_{\omega,x}\times w_{i,x,in,j}+b_{i,x,h,j})\times w_{i,x,o,j})+b_{i,x,o}).
\label{komega}
 \end{equation}

\begin{figure}[]
\centering
\subfigure[The architecture of the ANN (ANN\textsubscript{r,x}) to calculate the real part of the refractive index of material $x$.] {\label{fig:ANNr}{\includegraphics[width=0.46\textwidth]{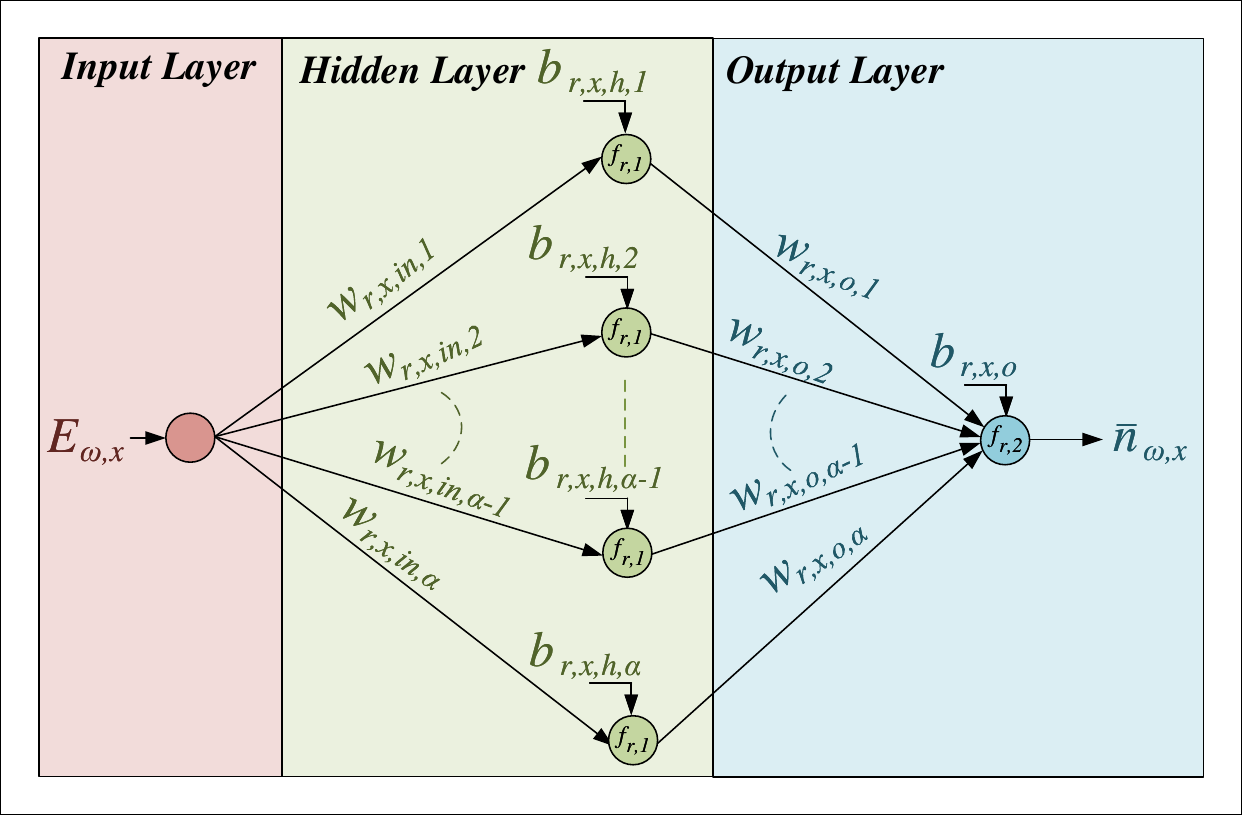}}}
\subfigure[The architecture of the ANN (ANN\textsubscript{i,x}) to calculate the imaginary part of the refractive index of material $x$.]
{\label{fig:ANNi}{\includegraphics[width=0.46\textwidth]{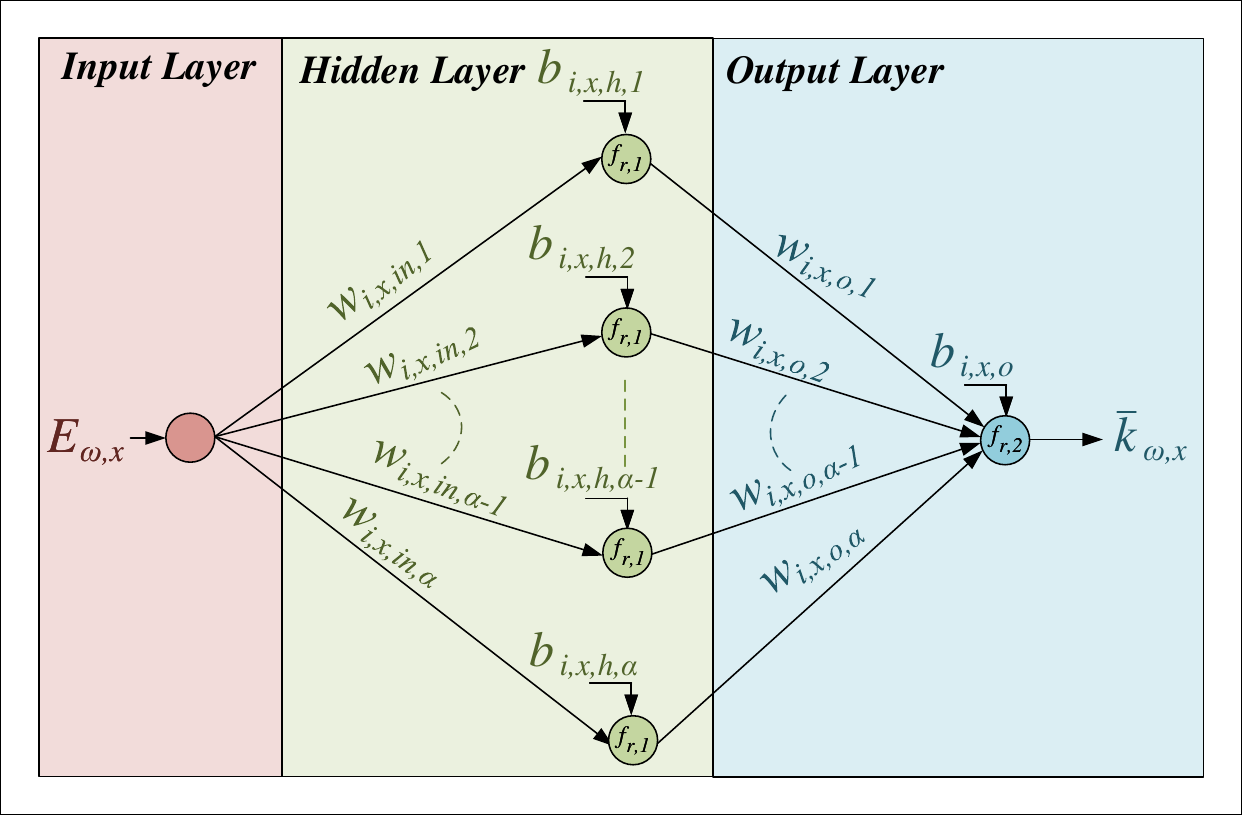}}}
\caption{The representation of: (a) the architecture of ANN\textsubscript{r,x} to calculate $n_{\omega,x}$ using weighting factors and the bias coefficients, and (b) the architecture of ANN\textsubscript{i,x} to calculate $k_{\omega,x}$ using weighting factors and the bias coefficients.}
\label{fig:ANN}
\end{figure}

 In \eqref{nomega}, $f_{r,1}$ and $f_{r,2}$ are the activation functions of the hidden layer and the output layer of ANN\textsubscript{r,x}. In addition, in \eqref{komega} $f_{i,1}$ and $f_{i,2}$ represent the activation functions of the hidden layer and the output layer of ANN\textsubscript{i,x}, respectively. For more clarification, Fig. \ref{fig:ANN} illustrates the architecture of the ANNs and the way of the calculation of the output of the ANNs. Based on Fig. \ref{fig:ANNr} and Fig. \ref{fig:ANNi}, the input data (the photon energy) is used as the input of ANN\textsubscript{r,x} and ANN\textsubscript{i,x}, and that is propagated by the hidden layer to the output layer to calculate the real part and the imaginary part of the refractive index. The task of the training is an optimization strategy to calculate the optimized values of the weighting coefficients and the bias factors of ANN\textsubscript{r,x}and ANN\textsubscript{i,x} to have the maximum level of accuracy in the estimation of $n_{\omega,x}$ and $k_{\omega,x}$ using \eqref{nomega} and \eqref{komega}, respectively. In the next Subsection, the structure of the transferred knowledge of ANNs related to copper will be presented.

\subsection{Knowledge Preparation}

The parameters of ANNs for material $x$ (i.e., ANN\textsubscript{r,x} and ANN\textsubscript{i,x}) can be defined as follows:

    \begin{equation}
\Upsilon_{r,x}=\begin{bmatrix}
W_{r,x,in} & W_{r,x,o} & B_{r,x,h}  & B_{r,x,o}
\end{bmatrix},
\label{v110}
 \end{equation}

 and,

     \begin{equation}
\Upsilon_{i,x}=\begin{bmatrix}
W_{i,x,in} & W_{i,x,o} & B_{i,x,h}  & B_{i,x,o}
\end{bmatrix}.
\label{v111}
 \end{equation}

 Therefore, the vector representation of parameters that are related to ANNs for metal $x$ is as follows:

     \begin{equation}
\Upsilon_{x}=\begin{bmatrix}
\Upsilon_{r,x}  & \Upsilon_{i,x}
\end{bmatrix},
\label{v112}
 \end{equation}

The knowledge of the neural networks for a material such as $g$ ($KN_{g}$) which will be transferred to train new ANNs for other metals is as follows:

     \begin{equation}
KN_{g}=\begin{Bmatrix}
\Upsilon_{x} & h_n & A_{g}
\end{Bmatrix}.
\label{v113}
 \end{equation}

In \eqref{v113}, $\Upsilon_{g}$ can be constructed using \eqref{v112} ($x==g$), $h_n$ is the number of the hidden layers, and $A_{g}$ represents the number of the neurons of the hidden layers. As mentioned before, in this study, only one hidden layer with $\alpha$ neurons is used. Therefore, $h_n=1$, and $A_{g}=\alpha$. In the next Subsection, the prepared knowledge will be used for the transfer learning-based strategy to tune the new ANNs, which will be used for another material.

\begin{figure}
	\centering
	\includegraphics[width=1\linewidth] {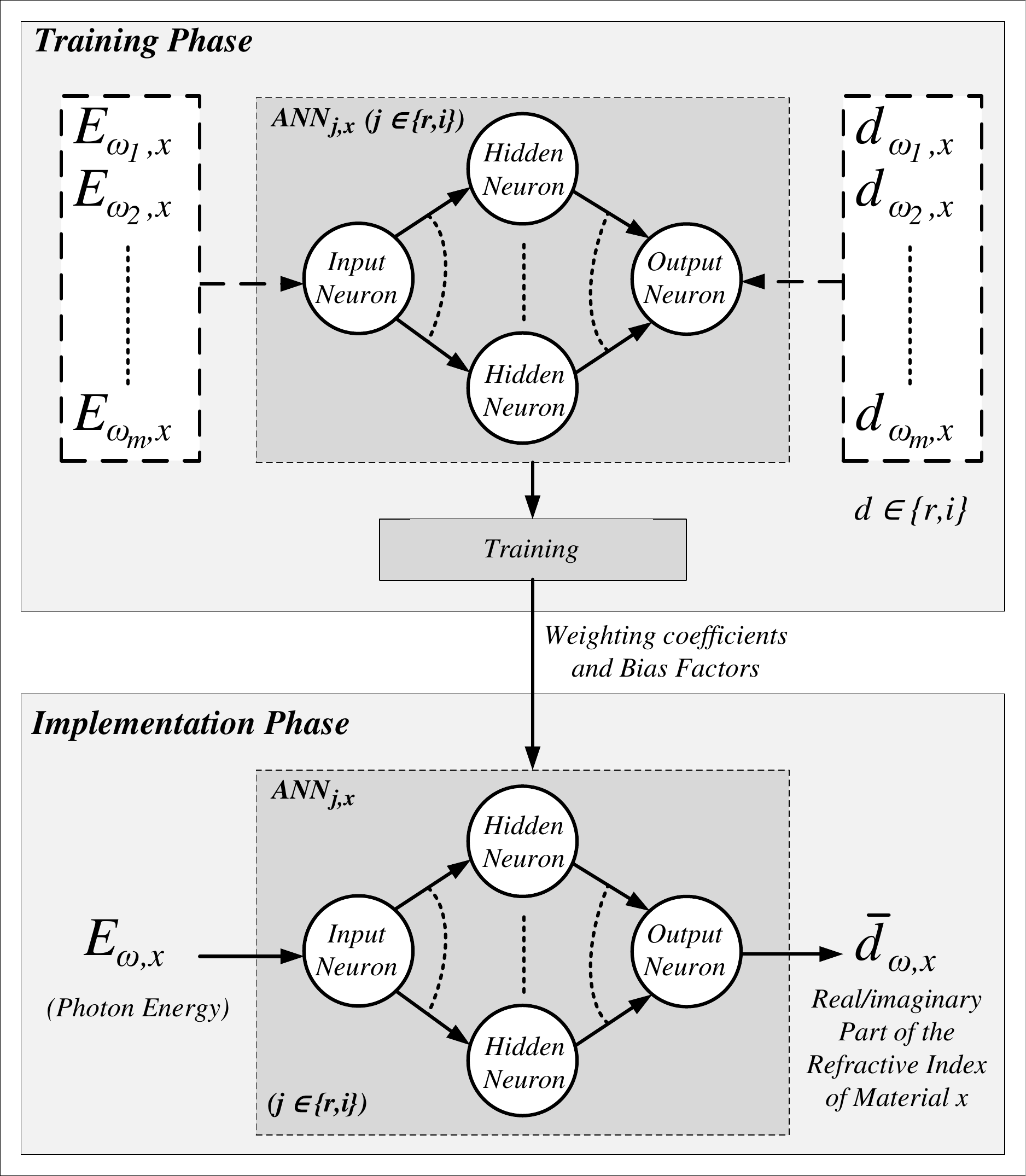}
	\caption{The training and the iplementation phase of ANNs to calculate the real and the imaginary parts of the refractive index of material $x$, where, $r==n$ and $i==k$.}
	\label{fig:Transfer_Learning}
\end{figure}

\subsection{Knowledge Implementation for the Transfer Learning-based strategy}

The variables that should be considered for the training the ANNs for a material $s$ is as follows:

     \begin{equation}
VR_{r,x}=\Upsilon_{r,x} \quad \quad and \quad \quad x\in \begin{Bmatrix}
s
\end{Bmatrix},
\label{v114}
 \end{equation}


and,

      \begin{equation}
VR_{i,x}=\Upsilon_{i,x} \quad \quad and \quad \quad x\in \begin{Bmatrix}
s
\end{Bmatrix}.
\label{v115}
 \end{equation}

 Where, $VR_{r,x}$ and  $VR_{i,x}$ include the variables for the training of ANN\textsubscript{r,x} and ANN\textsubscript{i,x}, respectively. In addition to the variables, a cost function should be considered to represent the training accuracy and performance. The cost function can be defined in a way to satisfy the minimum value of the error related to the training dataset. In other words, if the dataset for the training of ANN\textsubscript{r,x} is defined as follows:

      \begin{equation}
TrData_{r,x}=\begin{bmatrix}
Input_{r,x} & Output_{r,x}
\end{bmatrix},
\label{v116}
 \end{equation}
 where, $Input_{r,x}$ and $Output_{r,x}$ are the dataset of the input and the output of the ANN.
       \begin{equation}
Input_{r,x}=\begin{bmatrix}
E_{\omega_1,x} & E_{\omega_2,x} & \cdots & E_{\omega_{m-1},x}  & E_{\omega_m,x}
\end{bmatrix}^T,
\label{v117}
 \end{equation}
 and,
       \begin{equation}
Output_{r,x}=\begin{bmatrix}
n_{\omega_1,x} & n_{\omega_2,x} & \cdots & n_{\omega_{m-1},x}  & n_{\omega_m,x}
\end{bmatrix}^T.
\label{v118}
 \end{equation}

 \begin{figure*}
	\centering
	\includegraphics[width=1\linewidth] {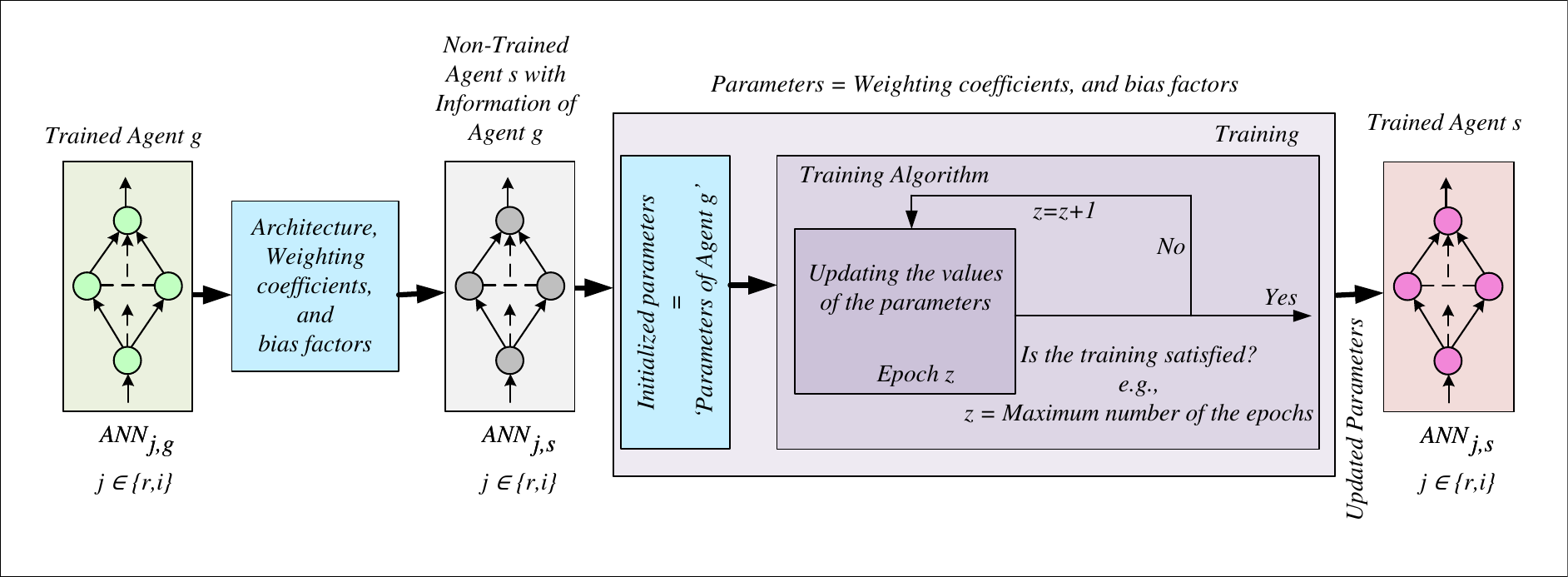}
	\caption{The implementation of transfer learning to transfer the knowledge of agent (material) $g$ to be used to train the ANNs for agent (material) $s$.}
	\label{fig:Transfer_Learning1}
\end{figure*}

In \eqref{v117}, $E_{\omega_j,x}$ is the photon energy at frequency $\omega_j$ related to material $x$, and $m$ is the number of samples in the training dataset. In addition, in \eqref{v118}, $n_{\omega_j,x}$ is the real value of the optical property of material $x$ at frequency $\omega_j$. As for ANN\textsubscript{r,x}, the dataset of the training of ANN\textsubscript{i,x} can be defined in a same way as follows:

      \begin{equation}
TrData_{i,x}=\begin{bmatrix}
Input_{i,x} & Output_{i,x}
\end{bmatrix},
\label{v119}
 \end{equation}
 where,
       \begin{equation}
Input_{i,x}=\begin{bmatrix}
E_{\omega_1,x} & E_{\omega_2,x} & \cdots & E_{\omega_{m-1},x}  & E_{\omega_m,x}
\end{bmatrix}^T,
\label{v120}
 \end{equation}
 and,
       \begin{equation}
Output_{i,x}=\begin{bmatrix}
k_{\omega_1,x} & k_{\omega_2,x} & \cdots & k_{\omega_{m-1},x}  & k_{\omega_m,x}
\end{bmatrix}^T.
\label{v121}
 \end{equation} 

In \eqref{v121}, $k_{\omega_j,x}$ is the imaginary value of the optical property of material $x$ at frequency $\omega_j$.

Based on the training dataset, the cost function for ANN\textsubscript{r,x} and ANN\textsubscript{i,x} can be defined as mean squared error.

       \begin{equation}
Cost_{r,x}=\frac{\sum_{j=1}^{m}(n_{\omega_j,x}-\overline{n}_{\omega_j,x})^2}{m},
\label{v122}
 \end{equation} 

       \begin{equation}
Cost_{i,x}=\frac{\sum_{j=1}^{m}(k_{\omega_j,x}-\overline{k}_{\omega_j,x})^2}{m}.
\label{v123}
 \end{equation} 

 Based on the defined cost function, the goal of the training process is to find the parameters of the networks, which can lead to the minimum values of the cost functions. Now, based on the definitions of the optimization variables and the cost functions, a training process can be implemented to tune the parameters of the networks. For the number of iterations, the values of the variables will be updated, i.e., adding some values at each iteration to the variables in a way that the updated variables make the value of the cost function smaller. The training requires initialization of optimization variables, and then updating the value of them. Finally, the values of the variables with the best cost function (i.e., the minimum value) will be selected to be used as the parameters of the neural network. In Fig. \ref{fig:Transfer_Learning}, a general overview for the training and the implementation phases related to an ANN to estimate the real/imaginary part of the refractive index of material $x$ is depicted. Based on Fig. \ref{fig:Transfer_Learning}, for the training phase, a dataset including samples of the input and the output of the system is prepared. Then, based on the architecture of the ANN, the unknown variables of the ANN are defined (the weighting coefficients and the bias factors). Subsequently, training process is done to calculate the optimal values of the unknown parameters deploying the dataset. Finally, the trained ANN is used in the implementation phase to calculate the output of the ANN using new inputs.

To have the knowledge of the trained ANNs for material $g$, the structures of the ANNs for material $x$ are the same as the structure of ANNs for $g$, i.e., they have the same hidden layers, the same numbers of neurons for the hidden layers, and same variables as the inputs and the outputs. In addition, the values of the weighting factors and bias coefficients of the ANN of material $g$ are assigned as the initial values of the variables.

  \begin{algorithm}
\caption{Optical Constants Estimation based on Transfer Learning}
\textbf{Step 1. Knowledge Preparation:} Preparation of the knowledge of an specific material like namely $g$ (i.e., $KN_{g}$) based on \eqref{v113} for transferring to train new ANNs for another material, i.e., namely material $s$.

\textbf{Step 2. Data Generation:} In this step, a dataset for training the $ANN_{l,x}$ ($l\in \begin{Bmatrix} r,i \end{Bmatrix}$, and $x\in \begin{Bmatrix} s \end{Bmatrix}$) is prepared. To increase the size of the dataset, a linear interpolation between samples can be considered. Then, based on the existing and the generated data, $TrData_{l,x}$ are created by \eqref{v116} and \eqref{v119}.

\textbf{Step 3. Cost Function Definition:} In this part, the cost function $Cost_{l,x}$ based on \eqref{v122} and \eqref{v123} is defined.

\textbf{Step 4. Knowledge Transfer to Training Units:} Here, the parameters of the ANN of material $g$ will be assigned to the initial values of the weighting coefficients and the bias factors of $ANN_{l,x}$ ($l\in \begin{Bmatrix} r,i \end{Bmatrix}$, and $x\in \begin{Bmatrix} s \end{Bmatrix}$).

\textbf{Step 5. Updating the Values of the Variables:} The values of the variables are updated using an iterative-based training strategy.

\textbf{Step 6. Selection the Variables with the Best Cost Function:} Select the variables with the best (minimum cost function) position and call it $B$.

\textbf{Step 7. Knowledge Preparation of the ANN:} Assigning $B_{l,x}$ to the parameters of $ANN_{l,x}$, i.e., $ \Upsilon_{l,x}= B_{l,x}$.

\textbf{Step 8. Adjusting the Weighting Coefficients and Bias Factors:} Finally, the weighting coefficients and the bias factors of $ANN_{l,x}$ can be find using \eqref{v110} and \eqref{v111}.

\label{Algstrategy}
\end{algorithm}

\subsection{Overview of the Strategy}

The transfer learning-based strategy is explained by Algorithm 1. In Algorithm 1, there are 8 steps. For the first step, the knowledge of well-trained ANNs for material $g$ is used to transfer to another unit to train the ANNs for material $s$. In addition, in Step 2, a linear interpolation between samples of the original dataset is used to increase the size of the dataset. In step 3, the cost function for the training process is defined. In addition, in Step 4, the knowledge that is prepared in Step 1 is used to initialize the weighting coefficients and the bias factors. Further, in Step 5, the values of the variables are updated. In addition, in Step 6, variables with the best performance are selected. To select the variables with the best performance, their values with the minimum cost function are selected. In Step 7, the knowledge for the targeted ANN is prepared based on the selected variables with the minimum value of the cost function. Then, in Step 8, the weighting coefficients and the bias factors of the targeted ANN are updated using the prepared knowledge in Step 7. In Fig. \ref{fig:Transfer_Learning1}, the general architecture of the transfer learning-based strategy (to transfer the knowledge of material $g$ to be used for the training of the ANNs for material $s$ to estimate the real and the imaginary parts of the refractive index) is depicted. The next parts, Section III and Section IV, provide the results of the trained ANNs to evaluate the performance of the transfer learning-based strategy for noble metal and semiconductor materials, respectively.

\section{Results I: Noble Metals}
\label{Sec:Results}

Here we discuss the results of the transfer learning-based strategy for noble metals. A dataset that is provided in \cite{PhysRevB.6.4370} is deployed as the original dataset with 49 samples. Then, a data generation technique is used to generate new samples from 0.640 eV to 6.596 eV, with a general energy step of $0.002$ eV. In addition, uniform random numbers between $-0.04$ to $0.04$ as well as between $-0.06$ to $0.06$ are added to the new samples of the real and imaginary optical constants of copper to play the role of the noise or error in the generated dataset. For the gold dataset, uniform random numbers between $-0.06$ to $0.06$ are added to the both real and imaginary parts of the refractive index. In the next part, the properties of the trained ANNs for the estimation of the optical constants of copper will be discussed.

\begin{figure}
	\centering
	\includegraphics[width=1.05\linewidth] {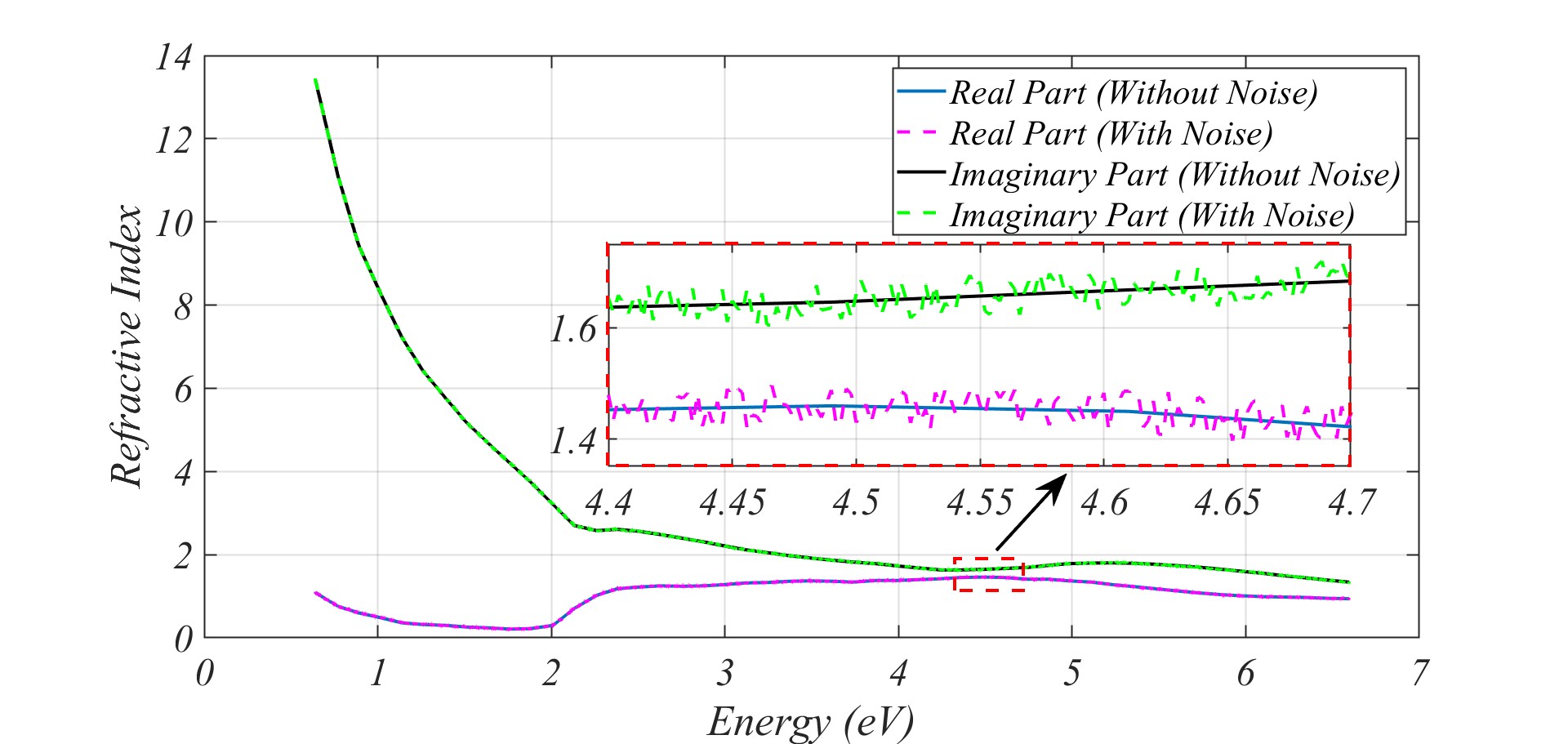}
	\caption{The real and the imaginary parts of the refractive index of Cu plotted as function of energy of light.}
	\label{fig:Dataset}
\end{figure}

\begin{figure}
	\centering
	\includegraphics[width=1.05\linewidth] {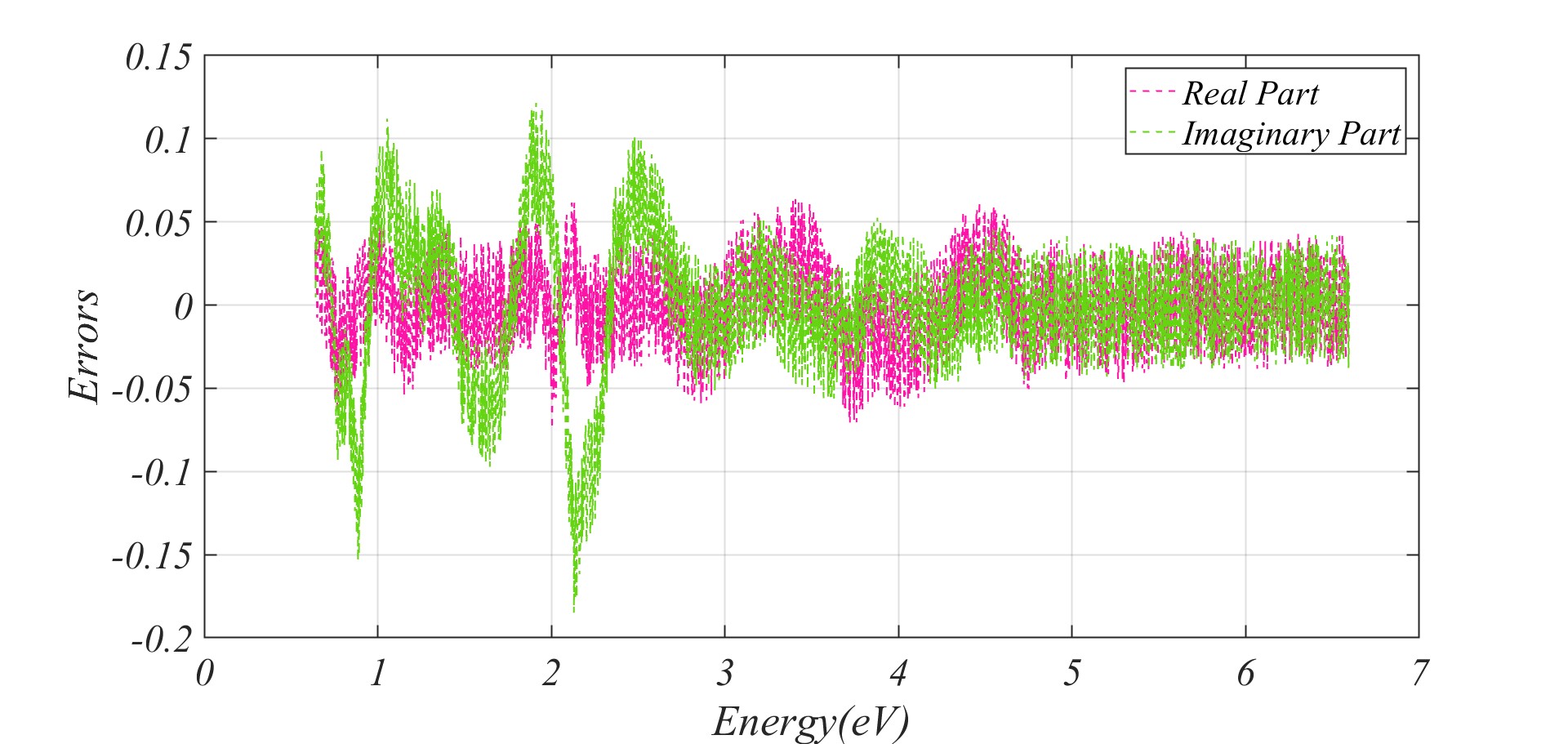}
	\caption{The error of the calculations related to ANN\textsubscript{r,Cu} and ANN\textsubscript{i,Cu} to estimate the real part and the imaginary part of the refractive index plotted as function of energy of light.}
	\label{fig:Train_imaginary_fit}
\end{figure}

\begin{figure}
	\centering
	\includegraphics[width=1\linewidth] {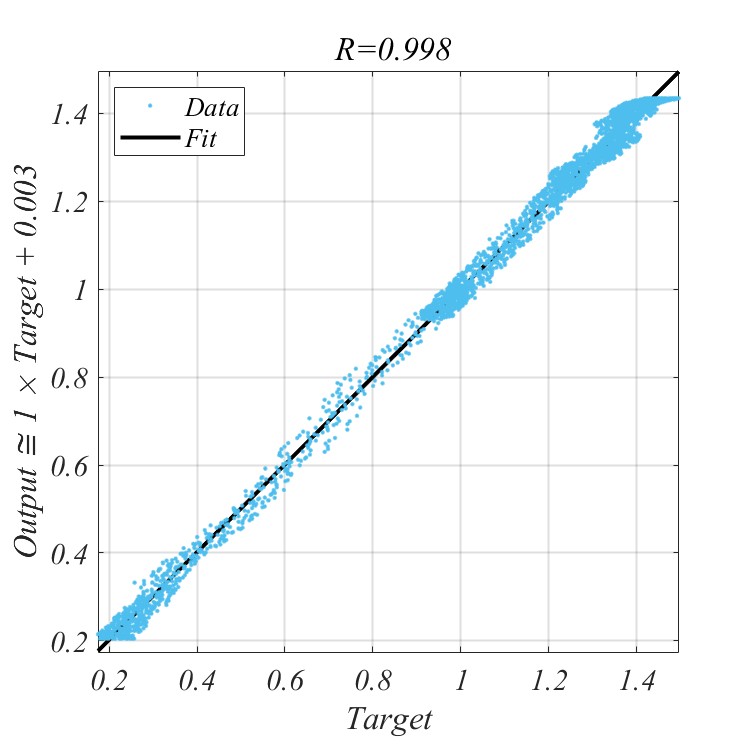}
	\caption{The regression plot related to the ANN to estimate the real part of the refractive index. The target is the implemented value of the real part of the refractive index, and the output is the estimated value of that, i.e., the output of ANN\textsubscript{r,Cu} (the regression coefficient is 0.998).}
	\label{fig:Train_real_regression}
\end{figure}

\begin{figure}
	\centering
	\includegraphics[width=1\linewidth] {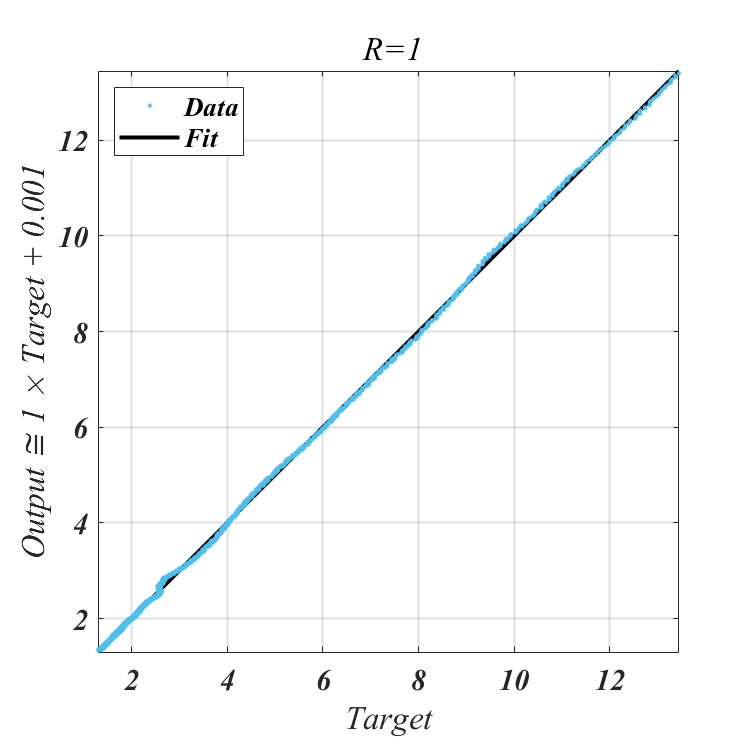}
	\caption{The regression plot related to the ANN to estimate the imaginary part of the refractive index. The target is the implemented value of the imaginary part of the refractive index, and the output is the estimated value of that, i.e., the output of ANN\textsubscript{i,Cu} (the regression coefficient is 1).}
	\label{fig:Train_imaginary_regression}
\end{figure}

\subsection{Information of the Trained ANNs for Cu}

Here, two ANNs to estimate the real and the imaginary parts of the refractive index of copper are trained. In Fig. \ref{fig:Dataset}, the real and the imaginary parts of the refractive index are depicted. It is important to note that, in Fig. \ref{fig:Dataset}, the generated dataset is shown, and the real and the imaginary parts of the refractive index with noise are deployed to train the ANNs. In addition, $70 \%$, $15 \%$, and  $15 \%$ of the dataset are used for the training, testing, and validation of the ANNs respectively, to have a large enough size of dataset for the training, examining the trained ANN after the training, and assessing the model during the training to avoid problems like overfitting. The selection of $70 \%$, $15 \%$, and  $15 \%$ of dataset allows us to use more than $\frac{2}{3} \approx 0.667$ of dataset to train the model and use a same portion of dataset (i.e., $\frac{100 \% -70\% }{2} = 15\%$) for the validation and testing to reach a well-trained network, which can be successfully examined. 

Furthermore, the difference between the real values and the estimated values of the real part and the imaginary part of the refractive index are shown in Fig. \ref{fig:Train_imaginary_fit}. Based on Fig. \ref{fig:Dataset}, altering the photon energy, the imaginary part of the refractive index of Cu has remarkable changes at lower energies and gradual variations at higher energies, which can lead to have larger errors at lower energies and smaller errors at higher energies. As it is shown by Fig. \ref{fig:Train_imaginary_fit}, the imaginary part has higher errors at lower energies. For a more accurate evaluation, parameters of $\theta_{r,x}$ and $\theta_{i,x}$ can be defined and used as follows:

 \begin{equation}
\theta_{r,x} =  \frac{\mu(\lvert n_{\omega_j,x} - \overline{n}_{\omega_j,x} \rvert)}{\mu(n_{\omega_j,x})},
\label{vthetaaaa111}
 \end{equation} 
and 
  \begin{equation}
\theta_{i,x} =  \frac{\mu(\lvert k_{\omega_j,x} - \overline{k}_{\omega_j,x} \rvert)}{\mu(k_{\omega_j,x})}.
\label{vthetaaaa1112}
 \end{equation} 
 
Where $\mu$ denotes the average value. Here, values of $\theta_{r,Cu}$ and $\theta_{i,Cu}$ are about 0.0207 and 0.0104, respectively. Also, in Fig. \ref{fig:Train_real_regression} and Fig. \ref{fig:Train_imaginary_regression}, the regression plots related to the neural networks for the real part and the imaginary part of the refractive index are illustrated. Based on Fig. \ref{fig:Train_real_regression} and Fig. \ref{fig:Train_imaginary_regression}, the regression coefficients for the estimation of the real and the imaginary parts of the refractive index are 0.998 and 1, respectively. Both values of the regression coefficients of the real and the imaginary parts are very close to 1 (approximately $\frac{|1-0.998|}{1} \times 100 = 0.2 \%$ and $\frac{|1-1|}{1} \times 100 = 0 \%$ deviation from 1 for the real and the imaginary parts), which shows the accuracy of the ANN models.

\subsection{Transfer Learning For Au}

This part will show the results of the trained ANNs for the estimation of the real part and the imaginary part of the refractive index of gold. The knowledge of ANN\textsubscript{r,Cu} and ANN\textsubscript{i,Cu} are used to train ANN\textsubscript{r,Au} and ANN\textsubscript{i,Au}, respectively. In addition, as for the condition related to the portion of dataset for Cu, the dataset for Au is also divided into three parts with values of $70\%$, $15\%$, and $15\%$ to train, validate, and test the trained ANNs, respectively. Also, in this part, ANN\textsubscript{r,Au} and ANN\textsubscript{i,Au} are trained using gradient descent backpropagation with a learning rate of $0.01$ and $0.001$. In addition, the number of the maximum epoch is 100000. Increasing the number of the maximum epoch can increase the computational burden of the training phase. However, increasing the maximum epoch can allow the model to converge to a more well-trained network. Therefore, 100000 is chosen as the maximum number of the epoch. As it will be shown later, based on the obtained results, with this number of epochs (i.e., 100000), the well-trained models are achieved with a very good performance.

\begin{figure}
	\centering
	\includegraphics[width=1.05\linewidth] {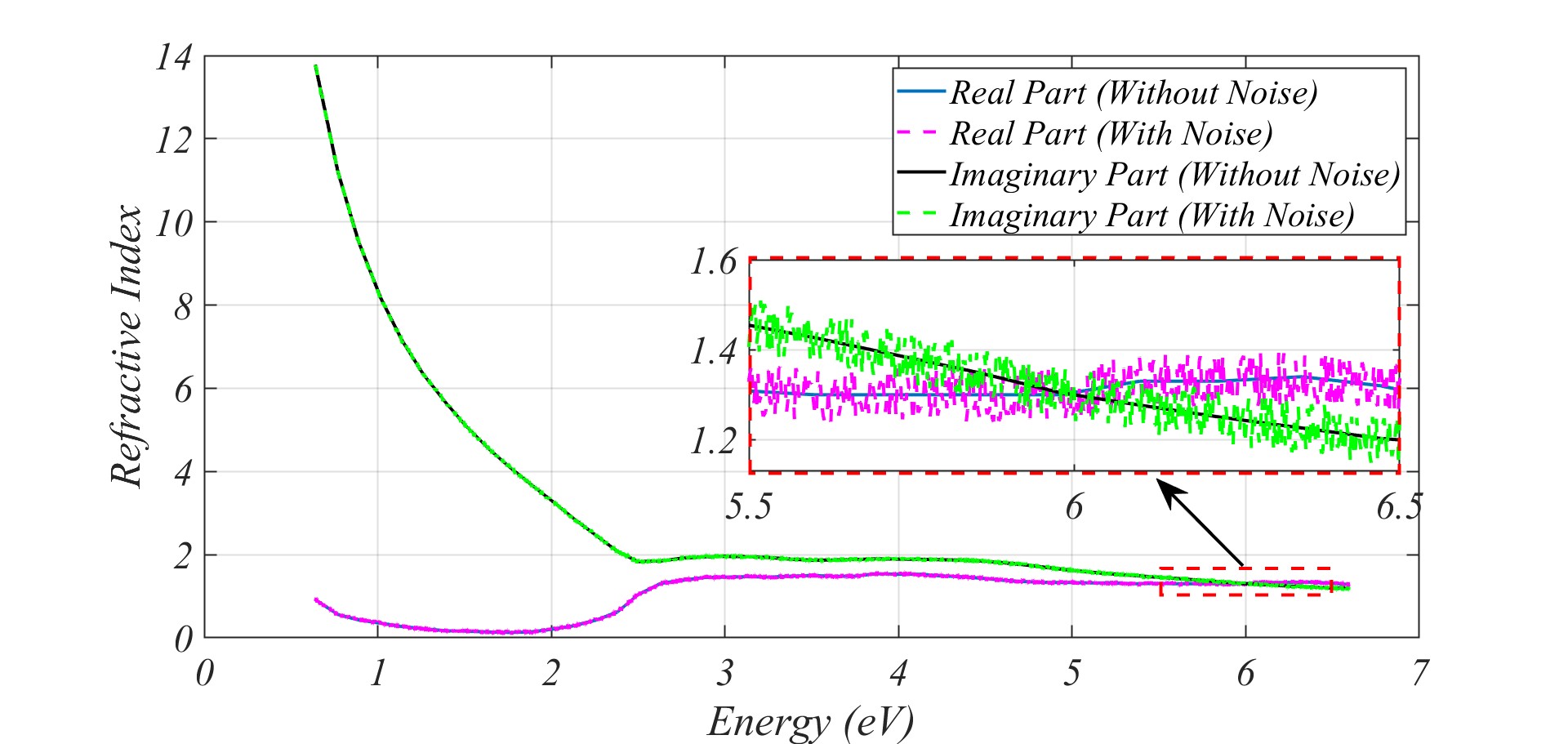}
	\caption{The real and the imaginary parts of the refractive index of Au plotted as function of energy of light.}
	\label{fig:Dataset_Gold}
\end{figure}

\begin{figure}
	\centering
	\includegraphics[width=1.05\linewidth] {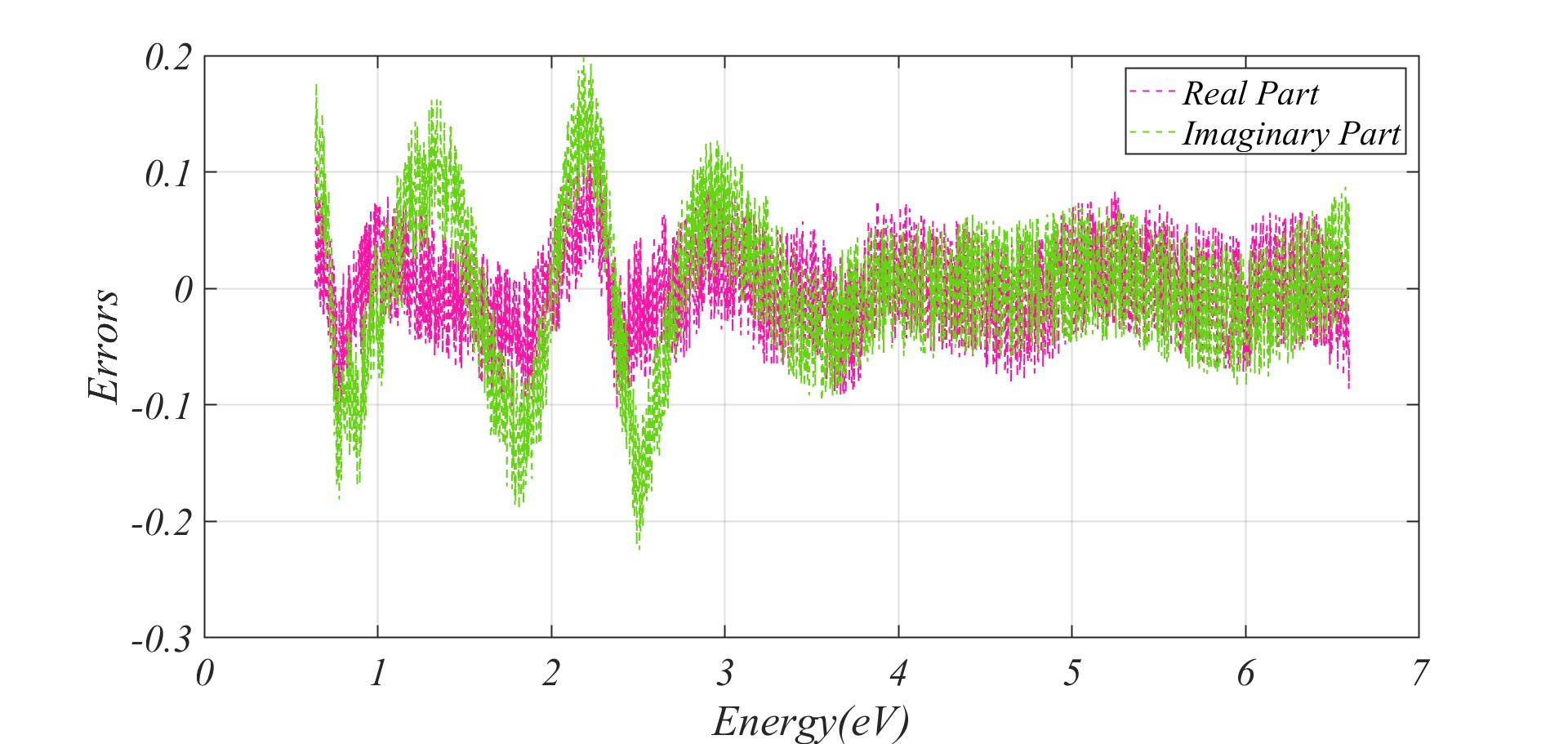}
	\caption{The error of the calculations related to ANN\textsubscript{r,Au} and ANN\textsubscript{i,Au} to estimate the real part and the imaginary part of the refractive index plotted as function of energy of light.}
	\label{fig:Train_imaginary_fit_gold}
\end{figure}

In Fig. \ref{fig:Dataset_Gold}, the real part and the imaginary part of refractive index of gold for various photon energies are shown. In this case, the noisy values of the refractive index are implemented to create the dataset for the training process. 
Furthermore, Fig. \ref{fig:Train_imaginary_fit_gold} depicts the error of the calculation related to ANN\textsubscript{r,Au} and ANN\textsubscript{i,Au}. Based on Fig. \ref{fig:Dataset_Gold}, the domain of the variations at lower energies is larger than changes at higher energies, which can lead to larger errors at lower energies, as shown in Fig. \ref{fig:Train_imaginary_fit_gold}. In addition, the values of $\theta_{r,Au}$ and $\theta_{i,Au}$ are 0.0319 and 0.0174, respectively.

\begin{figure}
	\centering
	\includegraphics[width=1\linewidth] {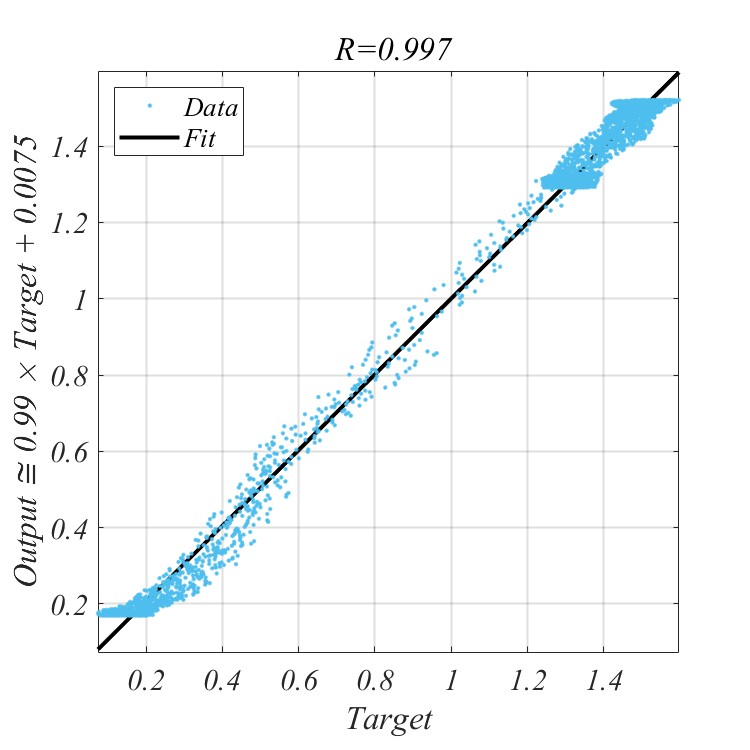}
	\caption{The regression plot related to the ANN to estimate the real part of the refractive index. The target is the implemented value of the real part of the refractive index, and the output is the estimated value of that, i.e., the output of ANN\textsubscript{r,Au} (the regression coefficient is 0.997).}
	\label{fig:Train_real_regression_gold}
\end{figure}

\begin{figure}
	\centering
	\includegraphics[width=1\linewidth] {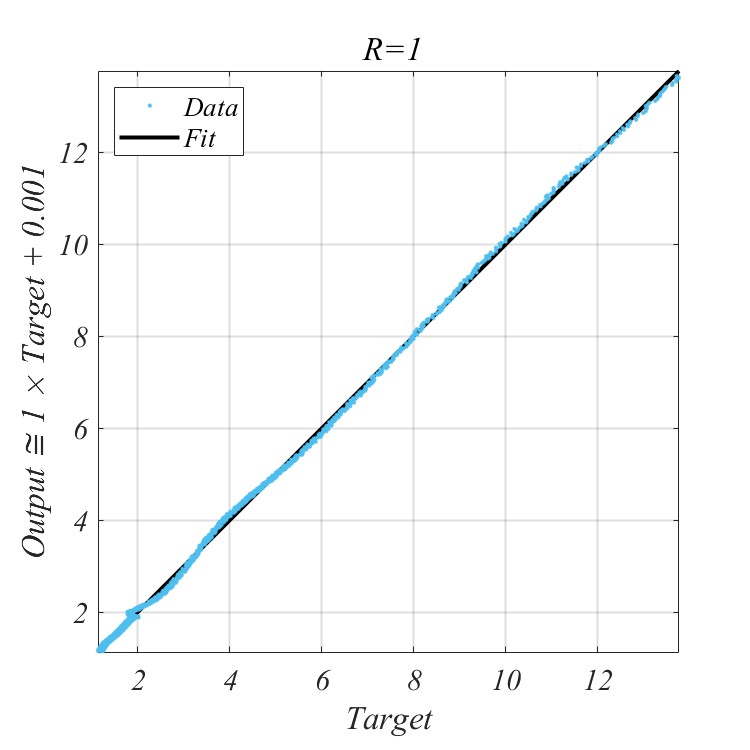}
	\caption{The regression plot related to the ANN to estimate the imaginary part of the refractive index. The target is the implemented value of the imaginary part of the refractive index, and the output is the estimated value of that, i.e., the output of ANN\textsubscript{i,Au} (the regression coefficient is 1).}
	\label{fig:Train_imaginary_regression_gold}
\end{figure}

In addition, in Fig. \ref{fig:Train_real_regression_gold} and Fig. \ref{fig:Train_imaginary_regression_gold}, the regression plot of ANN\textsubscript{r,Au} and ANN\textsubscript{i,Au} are shown. Based on Fig. \ref{fig:Train_real_regression_gold}, the line that can interpret the general functionality of ANN\textsubscript{r,Au} is $\overline{n}_{\omega,Au}=0.99 \times n_{\omega,Au}+0.0075$ and to the regression coefficient for ANN\textsubscript{r,Au} is 0.997, that is so close to 1. Furthermore, based on the Fig. \ref{fig:Train_imaginary_regression_gold}, the linear line to interpret and estimate the behaviour of ANN\textsubscript{i,Au} is $\overline{k}_{\omega,Au}=1 \times k_{\omega,Au}+0.001$ and the regression coefficient of ANN\textsubscript{i,Au} is 1. As discussed above, the interpreter linear lines of ANN\textsubscript{r,Au} and ANN\textsubscript{i,Au} are $\overline{n}_{\omega,Au}=0.99 \times n_{\omega,Au}+0.0075$ and $\overline{k}_{\omega,Au}=1 \times k_{\omega,Au}+0.001$, which are very close to lines of $\overline{n}_{\omega,Au}= n_{\omega,Au}$ and $\overline{k}_{\omega,Au}= k_{\omega,Au}$. In other words, the outputs of the ANNs are very close to the target values (i.e., based on the interpreter linear lines, it can be approximated that $\left| n_{\omega,Au} -  \overline{n}_{\omega,Au} \right| = \left| 0.01n_{\omega,Au} - 0.0075 \right|$ and $\left| k_{\omega,Au} -  \overline{k}_{\omega,Au} \right| =0.001$). Based on the obtained results, transfer learning can successfully support the ANNs to estimate the real part and the imaginary part of the refractive index of gold for different photon energies or consequently wavelengths of light.

\section{Results II: Semiconductor Materials}

To test the general usefulness and versatility of transfer learning-based approach, we examine  semiconductor materials using AI. Here, Si will be trained as the source of the knowledge. Then, the knowledge of the trained ANN for Si will be transferred to other materials, i.e., Ge, GaAs, GaP, GaSb, InP, and InAs. Then, based on the transferred knowledge, the ANNs will be trained for the mentioned semiconductor materials. It is important to note that, the datasets which are provided in  \cite{PhysRevB.27.985} are used. Here, for the dataset of Si, the data generation technique is used to increase the size of the dataset including more samples with a general energy step of 0.002 eV. As it will be shown by the captured results, this energy step allows having an enough large dataset to achieve reasonable and accurate results. Furthermore, for each dataset related to each semiconductor, uniform random numbers between -0.05 to 0.05 are added to the new samples of the real and imaginary parts of the refractive number to play the role of the noise on the dataset. Firstly, the trained ANN for Si will be presented. Then, the transfer learning-based strategy will be examined for other mentioned semiconductor materials.

\subsection{Information of the Trained ANNs for Si}

\begin{figure}
	\centering
	\includegraphics[width=1.05\linewidth] {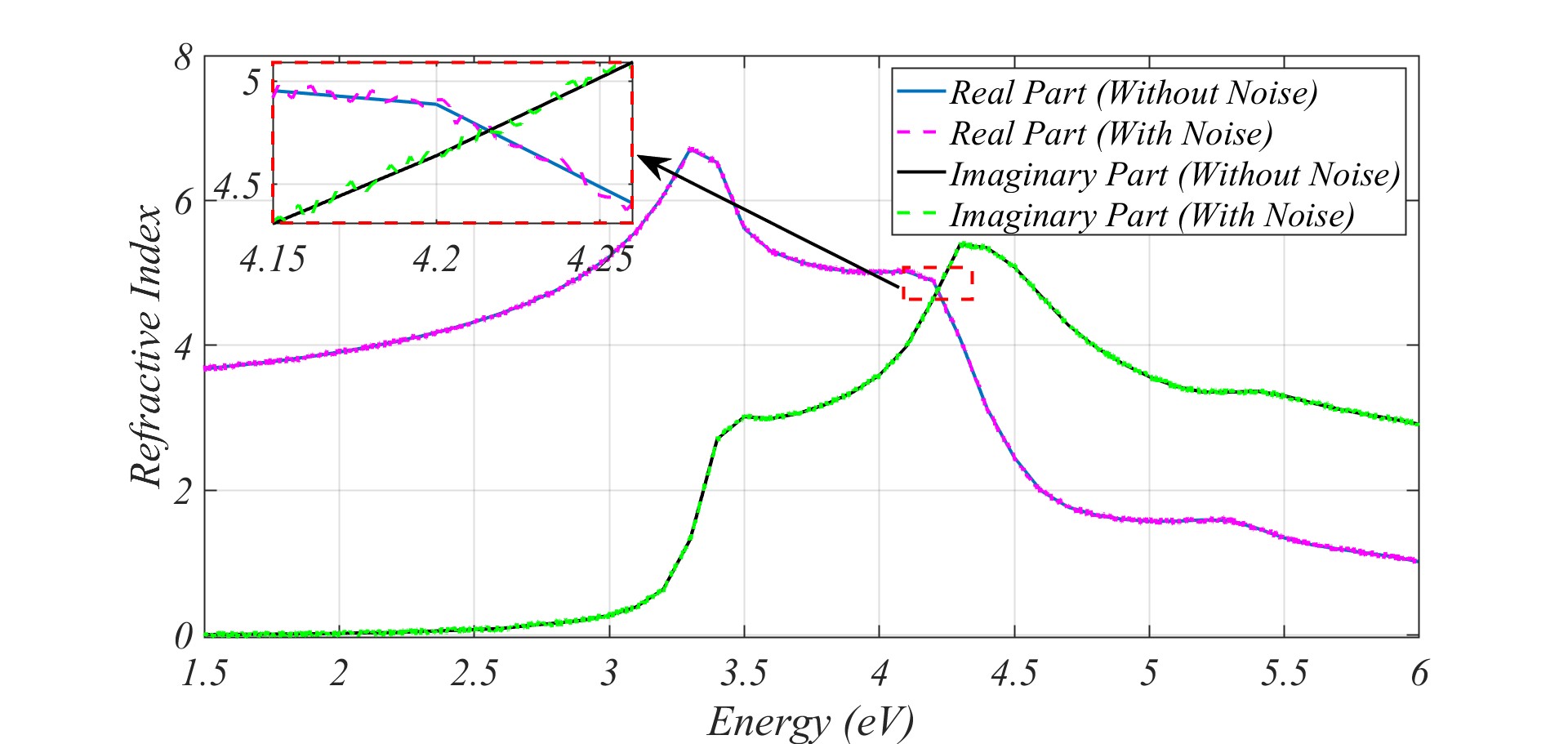}
	\caption{The real and the imaginary parts of the refractive index of Si plotted as function of energy of light.}
	\label{fig:Dataset_Si}
\end{figure}

ANNs are trained for the real and imaginary parts of the refractive index of Si. In Fig. \ref{fig:Dataset_Si}, the real and imaginary parts of the refractive index of Si are shown, with and without noise. \ref{fig:Dataset_Si}. 

\begin{figure}
	\centering
	\includegraphics[width=1.05\linewidth] {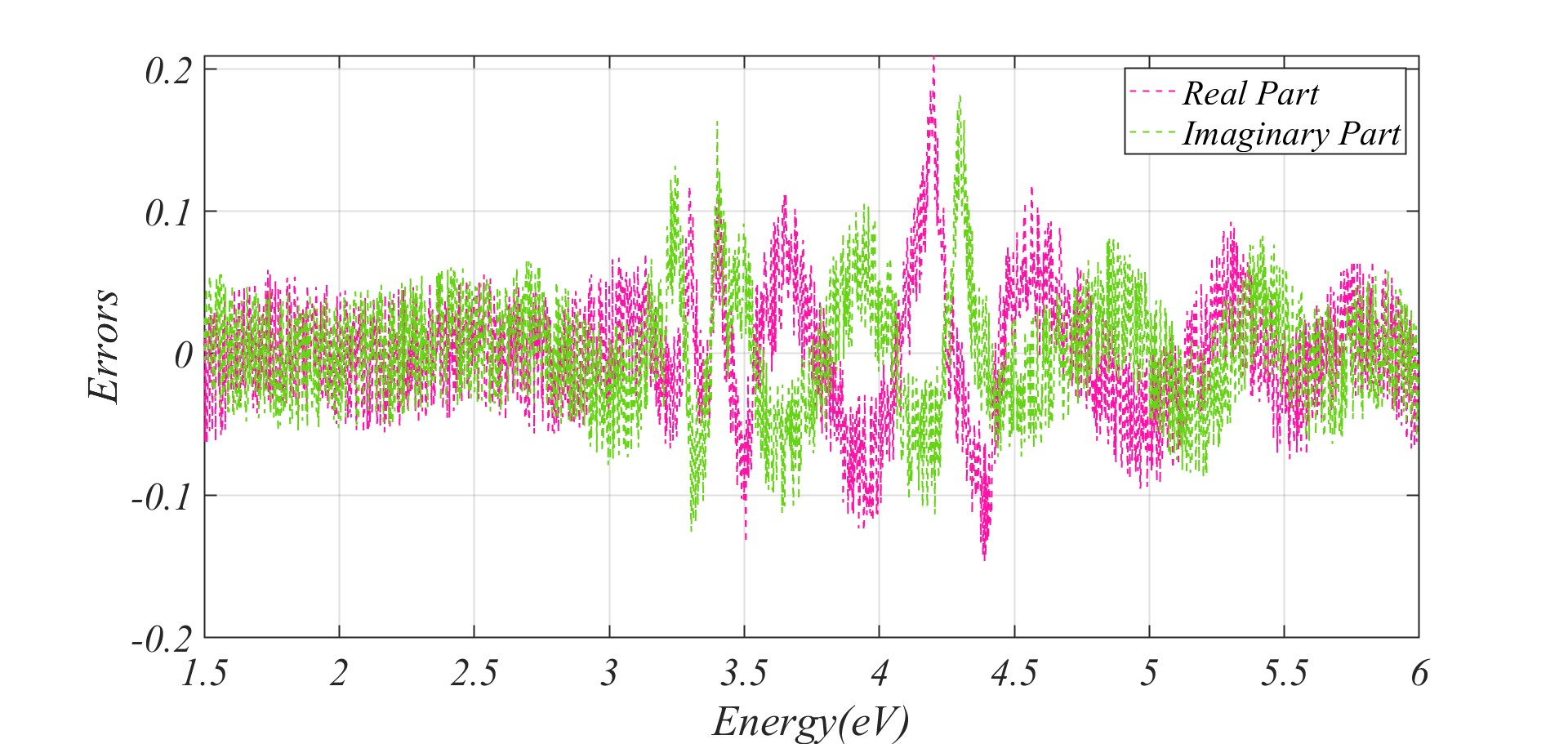}
	\caption{The error of the calculations related to ANN\textsubscript{r,Si} and ANN\textsubscript{i,Si} to estimate the real part and the imaginary part of the refractive index plotted as function of energy of light.}
	\label{fig:Train_imaginary_fit_Si}
\end{figure}

\begin{figure}
	\centering
	\includegraphics[width=1\linewidth] {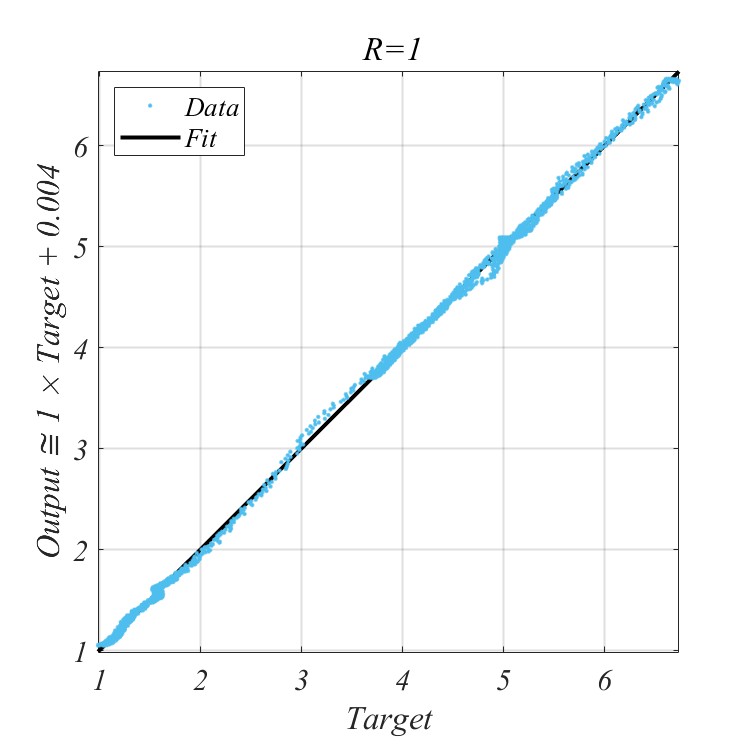}
	\caption{The regression plot related to the ANN to estimate the real part of the refractive index. The target is the implemented value of the real part of the refractive index, and the output is the estimated value of that, i.e., the output of ANN\textsubscript{r,Si} (the regression coefficient is 1).}
	\label{fig:Train_real_regression_Si}
\end{figure}

\begin{figure}
	\centering
	\includegraphics[width=1\linewidth] {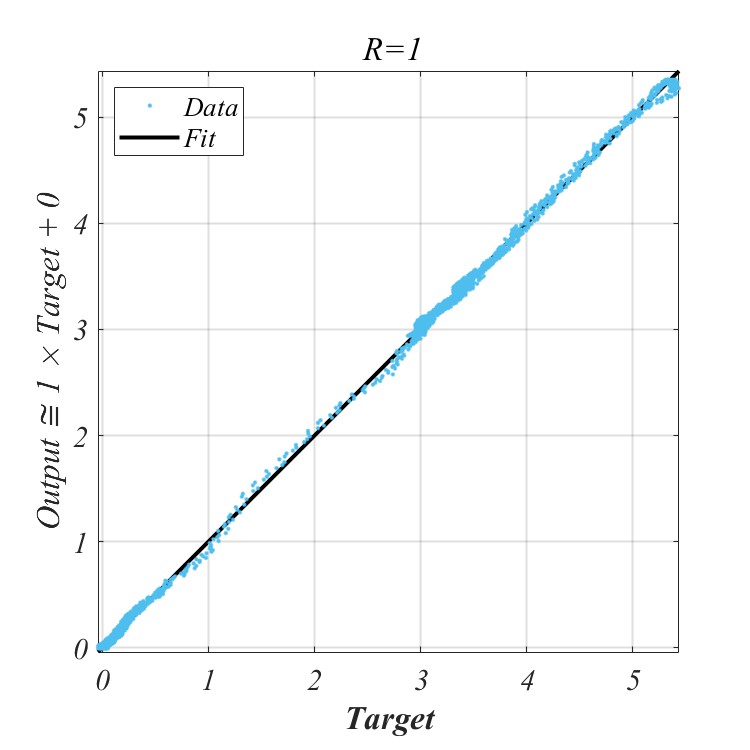}
	\caption{The regression plot related to the ANN to estimate the imaginary part of the refractive index. The target is the implemented value of the imaginary part of the refractive index, and the output is the estimated value of that, i.e., the output of ANN\textsubscript{i,Si} (the regression coefficient is 1).}
	\label{fig:Train_imaginary_regression_Si}
\end{figure}

Furthermore, the error of the calculations of the ANNs for the real part and the imaginary part of the refractive index is depicted in Fig. \ref{fig:Train_imaginary_fit_Si}. For this case study, based on \eqref{vthetaaaa111} and \eqref{vthetaaaa1112}, the values of $\theta_{r,Si}$ and $\theta_{i,Si}$ are 0.0099 and 0.0152, respectively. Finally, the regression plots of ANN\textsubscript{r,Si} and ANN\textsubscript{i,Si} are provided to have a clearer understanding of the performances of the ANNs to calculate the real and the imaginary parts of the refractive index of Si. The regression plots related to ANN\textsubscript{r,Si} and ANN\textsubscript{i,Si} are illustrated by Fig. \ref{fig:Train_real_regression_Si} and Fig. \ref{fig:Train_imaginary_regression_Si}, respectively. Based on the captured results, the regression coefficients for the real and the imaginary parts of the refractive index of Si are 1.

\begin{figure}[!t]
    \centering
    \subfigure[ The real and the imaginary parts of the refractive index of Ge plotted as function of energy of light.]{%
        \includegraphics[width=0.48\textwidth]{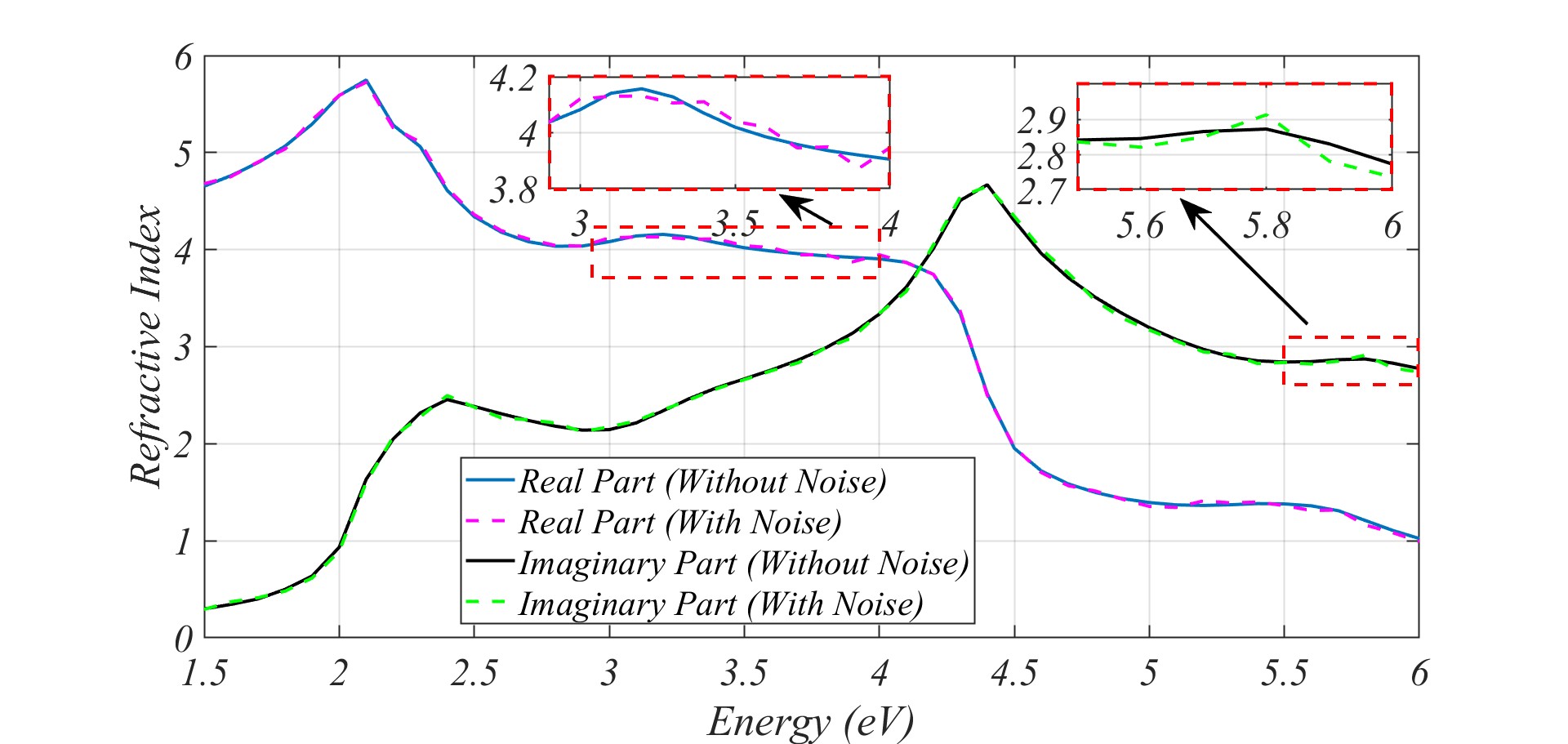}
        \label{fig:1Ge}
    }
    \subfigure[ The real and the imaginary parts of the refractive index of GaAs plotted as function of energy of light.]{%
        \includegraphics[width=0.48\textwidth]{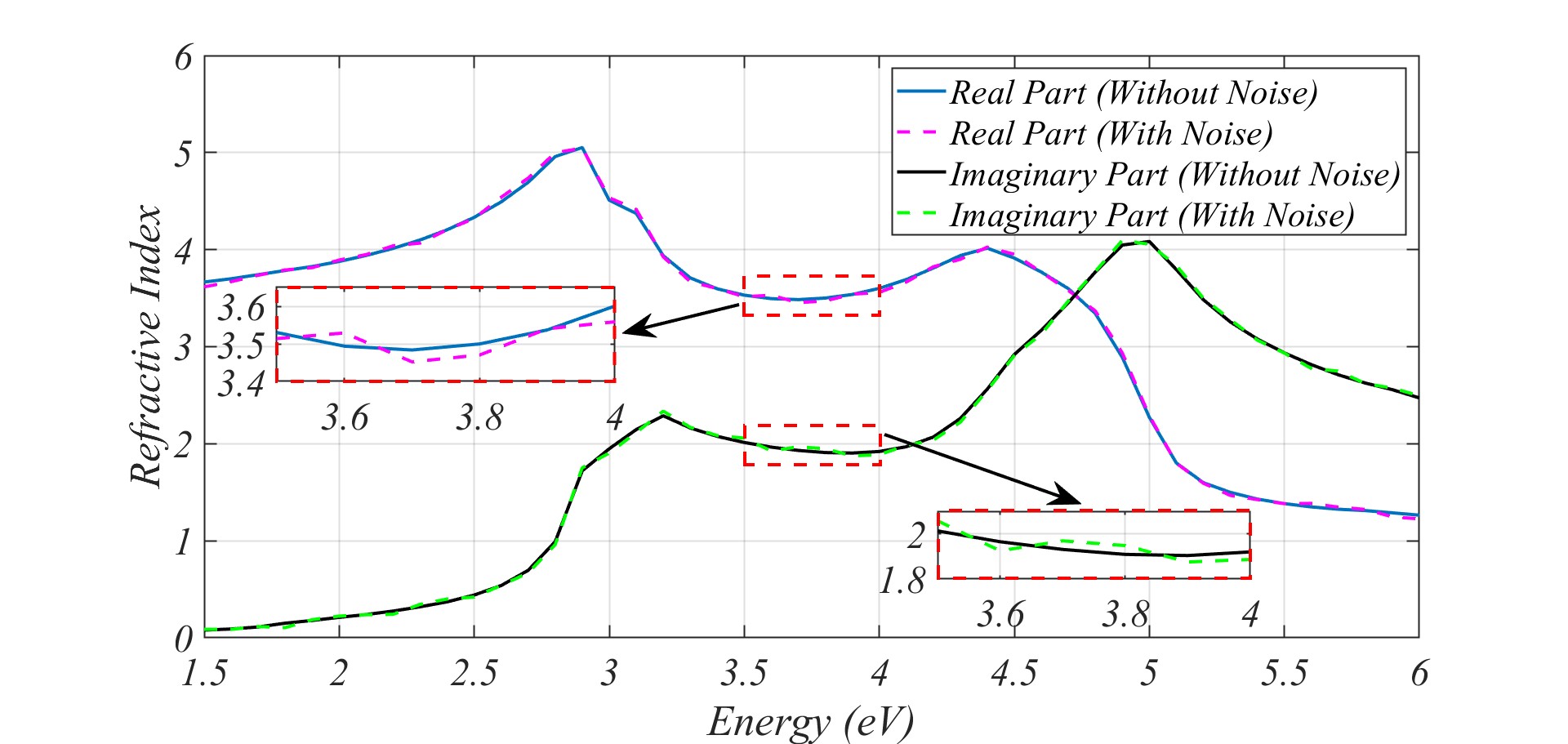}
        \label{fig:1GaAs}
    }
    \subfigure[ The real and the imaginary parts of the refractive index of InP plotted as function of energy of light.]{%
        \includegraphics[width=0.48\textwidth]{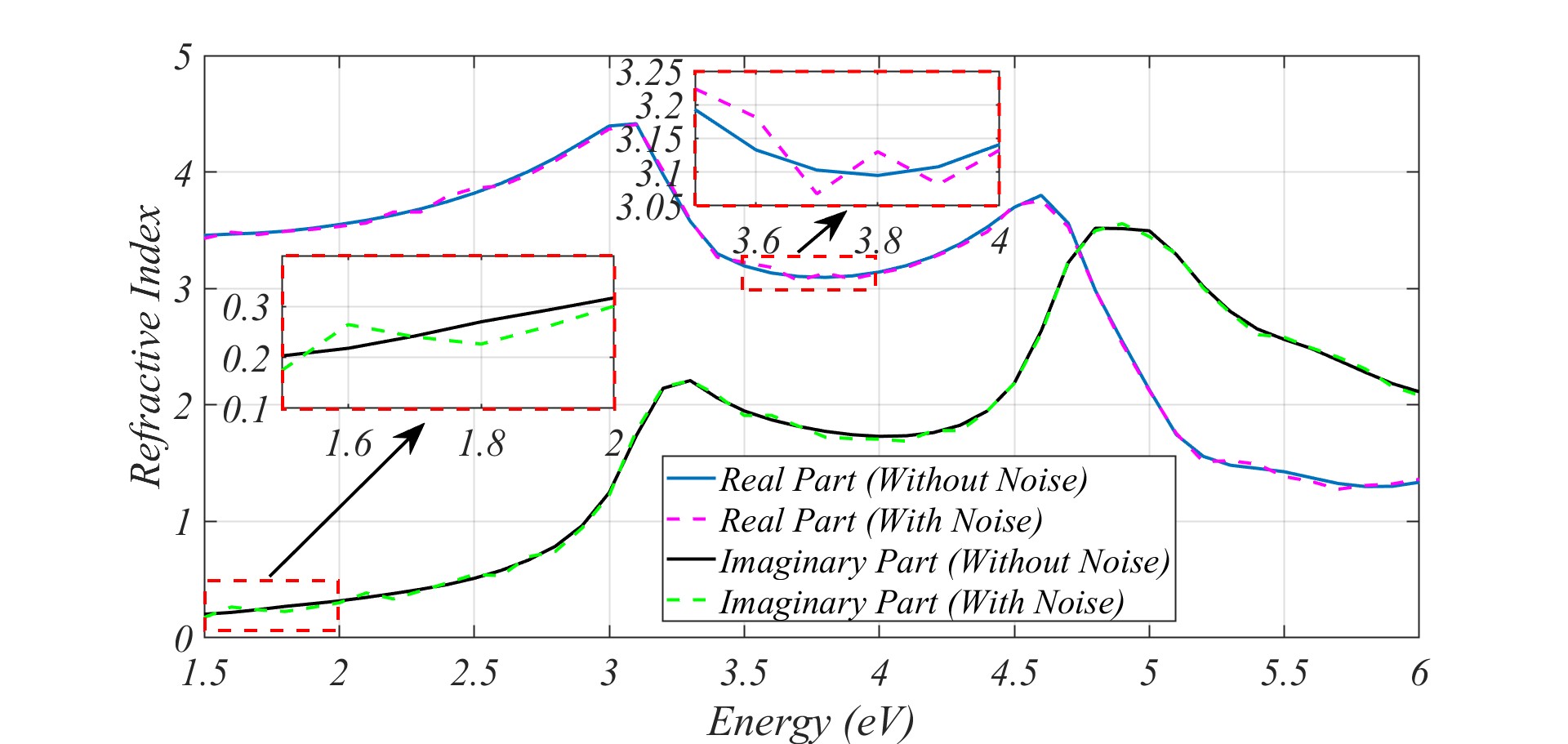}
        \label{fig:1InP}
    }
    \caption{The dataset related to the real part and the imaginary part of the refractive index for Ge, GaAs, and InP plotted as function of energy of light. In Fig. \ref{fig:1Ge}, Fig. \ref{fig:1GaAs}, and Fig. \ref{fig:1InP}, the dataset with and without the uniform noise (or the measurement error) for Ge, GaAs, and InP are depicted, respectively.}
    \label{fig:wholefigure123}
\end{figure}

\begin{figure*}[!t]
    \centering
    \subfigure[The regression plot related to the ANN to estimate the real part of the refractive index. The target is the implemented value of the real part of the refractive index, and the output is the estimated value of that, i.e., the output of ANN\textsubscript{r,Ge}.]{%
        \includegraphics[width=0.32\textwidth]{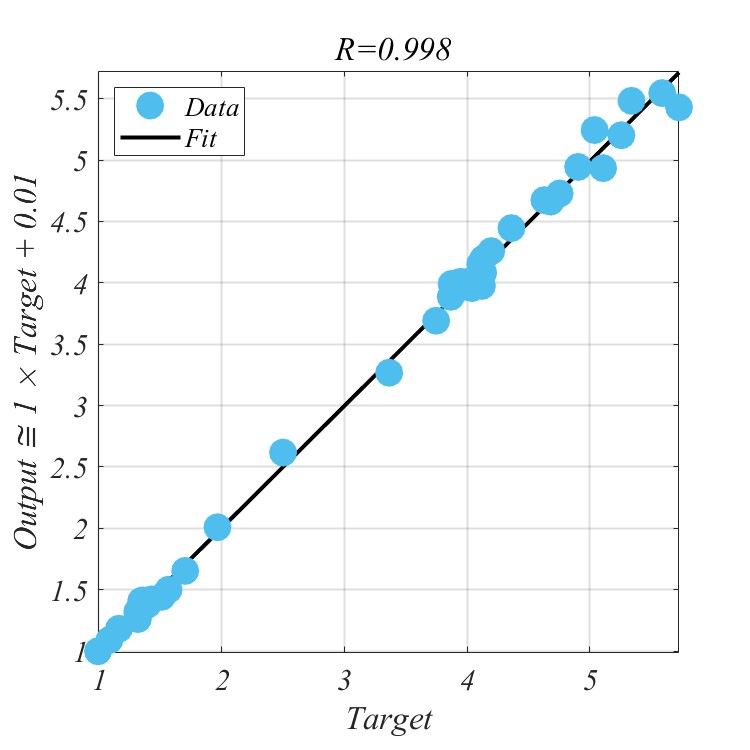}
        \label{fig:2Gereal}
    }
    \hfill
    \subfigure[The regression plot related to the ANN to estimate the real part of the refractive index. The target is the implemented value of the real part of the refractive index, and the output is the estimated value of that, i.e., the output of ANN\textsubscript{r,GaAs}.]{%
        \includegraphics[width=0.32\textwidth]{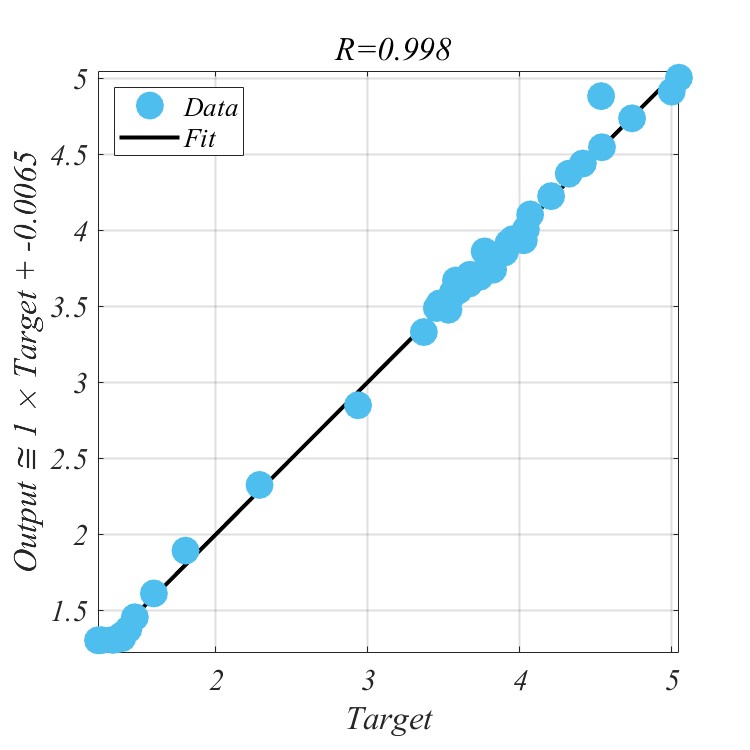}
        \label{fig:2GaAsreal}
    }
    \hfill
    \subfigure[The regression plot related to the ANN to estimate the real part of the refractive index. The target is the implemented value of the real part of the refractive index, and the output is the estimated value of that, i.e., the output of ANN\textsubscript{r,InP}.]{%
        \includegraphics[width=0.32\textwidth]{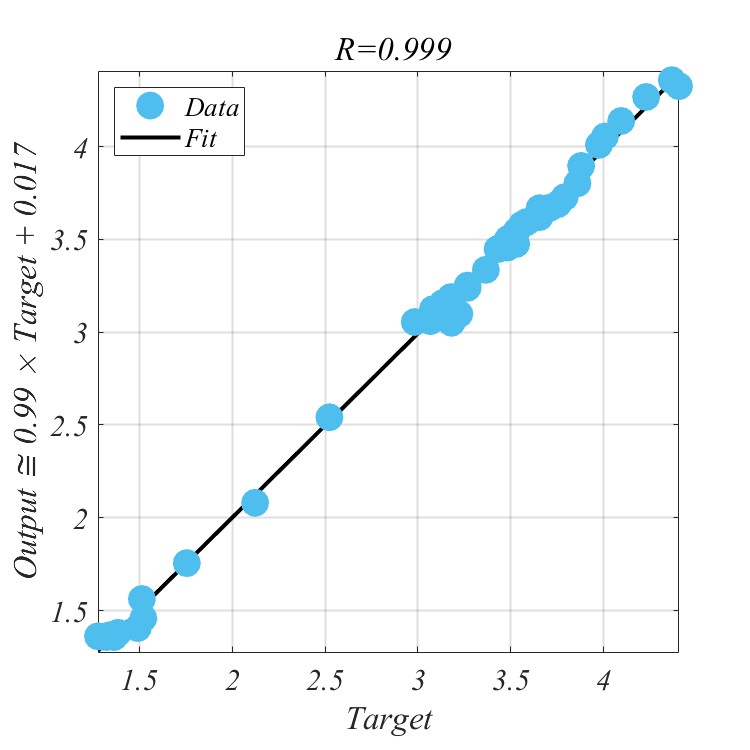}
        \label{fig:2InPreal}
    }
    \subfigure[The regression plot related to the ANN to estimate the imaginary part of the refractive index. The target is the implemented value of the imaginary part of the refractive index, and the output is the estimated value of that, i.e., the output of ANN\textsubscript{i,Ge}.]{%
        \includegraphics[width=0.32\textwidth]{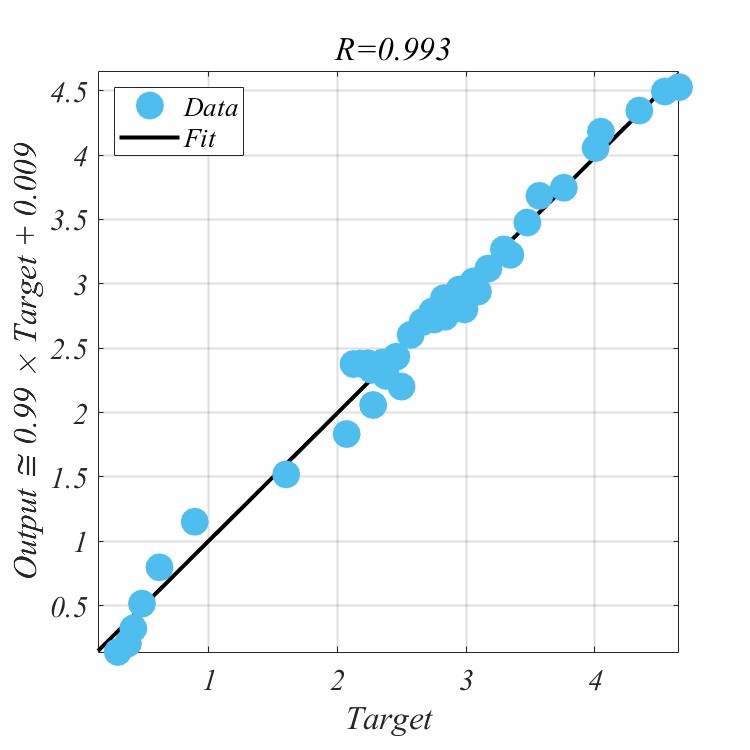}
        \label{fig:3Geimaginary}
    }
    \hfill
    \subfigure[The regression plot related to the ANN to estimate the imaginary part of the refractive index. The target is the implemented value of the imaginary part of the refractive index, and the output is the estimated value of that, i.e., the output of ANN\textsubscript{i,GaAs}.]{%
        \includegraphics[width=0.32\textwidth]{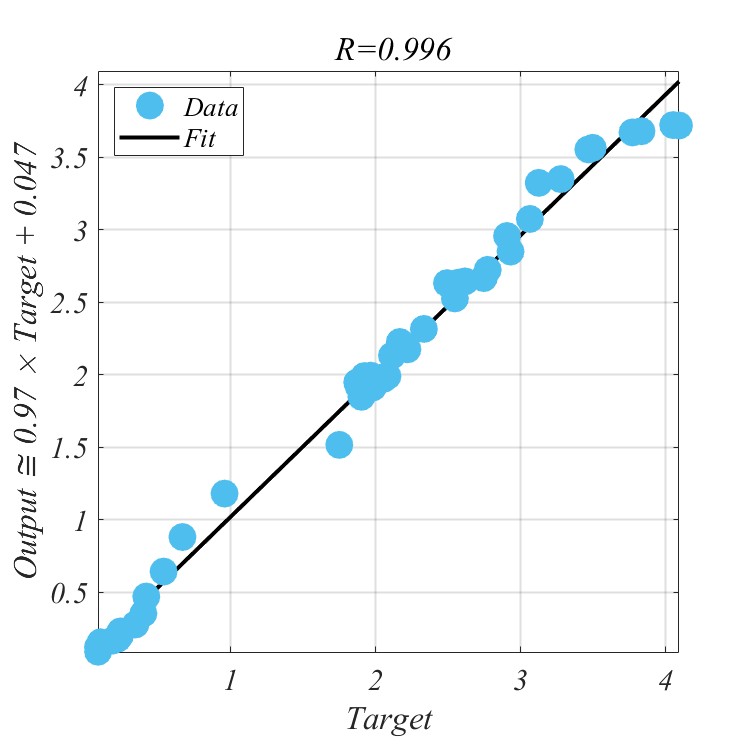}
        \label{fig:3GaAsimaginary}
    }
    \hfill
    \subfigure[The regression plot related to the ANN to estimate the imaginary part of the refractive index. The target is the implemented value of the imaginary part of the refractive index, and the output is the estimated value of that, i.e., the output of ANN\textsubscript{i,InP}.]{%
        \includegraphics[width=0.32\textwidth]{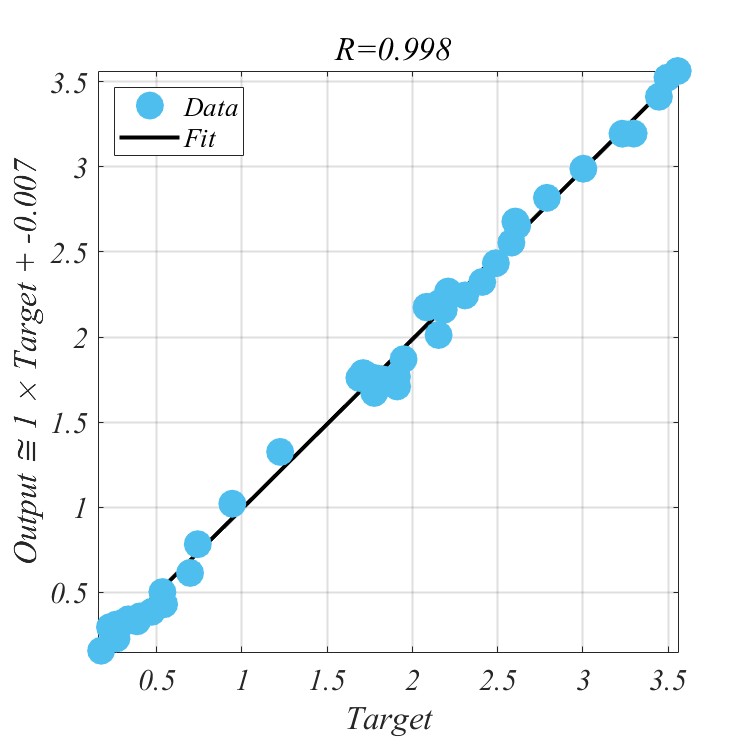}
        \label{fig:3InPimaginary}
    }
    \caption{The results of the neural networks to calculate the real part and the imaginary part of the refractive index for Ge, GaAs, and InP. In addition, Fig. \ref{fig:2Gereal}, Fig. \ref{fig:2GaAsreal}, and Fig. \ref{fig:2InPreal} illustrate the regression plot related to ANN\textsubscript{r,Ge}, ANN\textsubscript{r,GaAs}, and ANN\textsubscript{r,InP}, respectively. Further, the regression plot for ANN\textsubscript{i,Ge}, ANN\textsubscript{i,GaAs}, and ANN\textsubscript{i,InP}, are shown by Fig. \ref{fig:3Geimaginary}, Fig. \ref{fig:3GaAsimaginary}, and Fig. \ref{fig:3InPimaginary}, respectively.}
    \label{fig:wholefigure}
\end{figure*}

\subsection{Transfer Learning For Other semiconductor Materials, i.e., Ge, GaAs, and InP}

In this part, the knowledge of the trained ANNs for Si is transferred to be implemented to train ANNs to calculate the real part and the imaginary part of the refractive indices for Ge, GaAs, and InP. Here, the dataset includes the values of the refractive indices for energy levels from 1.5 eV to 6 eV with the energy steps of 0.1 eV. In addition, as mentioned before, uniform random data between $-0.05$ and $0.05$ are added to the original dataset to play the role of noise or the error of the measurements. Therefore, the dataset related to each semiconductor material includes 46 samples. In Fig. \ref{fig:1Ge}, Fig. \ref{fig:1GaAs}, and Fig. \ref{fig:1InP}, the real and the imaginary values of the refractive indices for the Ge, GaAs, and InP are illustrated. In addition, Fig. \ref{fig:2Gereal}, Fig. \ref{fig:2GaAsreal}, and Fig. \ref{fig:2InPreal} show the regression plots related to the real value of the refractive index for Ge (ANN\textsubscript{r,Ge}), GaAs (ANN\textsubscript{r,GaAs}), and InP (ANN\textsubscript{r,InP}), respectively. Based on Fig. \ref{fig:2Gereal}, Fig. \ref{fig:2GaAsreal}, and Fig. \ref{fig:2InPreal}, the regression coefficients related to ANN\textsubscript{r,Ge}, ANN\textsubscript{r,GaAs}, and ANN\textsubscript{r,InP} are $0.998$, $0.998$, and $0.999$, respectively.

\begin{table}[]
\centering
 \caption{Information of ANN\textsubscript{r,Ge}, ANN\textsubscript{r,GaAs}, and ANN\textsubscript{r,InP} with transfer learning.}
\begin{tabular}{c |c c}
Neural Network & Interpreter Line & Regression coefficient \\ \hline
\hline
ANN\textsubscript{r,Ge}             & $\overline{n}_{\omega,Ge}= n_{\omega,Ge}+0.01$             & 0.998                       \\
ANN\textsubscript{r,GaAs}           & $\overline{n}_{\omega,GaAs}= n_{\omega,GaAs}+0.0065$             & 0.998                       \\
ANN\textsubscript{r,InP}         & $\overline{n}_{\omega,InP}=0.99 n_{\omega,InP}+0.017$          & 0.999                      
\end{tabular}
\end{table}
\begin{table}[]
\centering
 \caption{Information of ANN\textsubscript{i,Ge}, ANN\textsubscript{i,GaAs}, and ANN\textsubscript{i,InP} with transfer learning.}
\begin{tabular}{c |c c}
Neural Network & Interpreter Line & $R$ coefficient \\ \hline
\hline
ANN\textsubscript{i,Ge}             & $\overline{k}_{\omega,Ge}= 0.99 k_{\omega,Ge}+0.009$             & 0.993                       \\
ANN\textsubscript{i,GaAs}           & $\overline{k}_{\omega,GaAs}= 0.97 k_{\omega,GaAs}+0.047$             & 0.996                       \\
ANN\textsubscript{i,InP}         & $\overline{k}_{\omega,InP}=0.99 k_{\omega,InP}+0.007$          & 0.998                      
\end{tabular}
\end{table}
\begin{table}[]
\centering
 \caption{More accurate value of the regression coefficient for Ge, GaAs, and InP.}
\begin{tabular}{c |c c}
Semiconductor & Real Part & Imaginary Part \\ \hline
\hline
Ge             & 0.99848             & 0.993                          \\
GaAs           & 0.99802               & 0.99576                         \\
InP         & 0.99879            & 0.9975                        
\end{tabular}
\end{table}
\begin{table}[]
\centering
 \caption{The value of $\theta_{r,x}$ and $\theta_{i,x}$ for Ge, GaAs, and InP.}
\begin{tabular}{c |c c}
Semiconductor & $\theta_{r,x}$ & $\theta_{i,x}$ \\ \hline
\hline
Ge             & 0.0180             & 0.03662                         \\
GaAs           & 0.0141               & 0.0427                         \\
InP         & 0.0121            & 0.0363                        
\end{tabular}
\end{table}
\begin{table}[]
\centering
 \caption{The validation performance for Ge, GaAs, and InP.}
\begin{tabular}{c |c c}
Semiconductor & Real Part & Imaginary Part \\ \hline
\hline
Ge             & 0.0093             & 0.0167                         \\
GaAs           & 0.0042               & 0.0376                         \\
InP         & 0.0036            & 0.0037                        
\end{tabular}
\end{table}

Furthermore, the regression plots of the neural networks for the imaginary parts of the refractive indices of Ge, GaAs, and InP are depicted in Fig. \ref{fig:3Geimaginary}, Fig. \ref{fig:3GaAsimaginary}, and Fig. \ref{fig:3InPimaginary}. Based on the Figures, the regression coefficients for calculating the imaginary parts of the refractive indices by ANN\textsubscript{i,Ge}, ANN\textsubscript{i,GaAs} , and ANN\textsubscript{i,InP} are $0.993$, $0.996$, and $0.998$, respectively. In Table I Table II, Table III, and Table IV, more information related to the neural networks can be found. Based on the interpreter lines of Table I and Table II, it can be approximated that $\left| n_{\omega,Ge} -  \overline{n}_{\omega,Ge} \right| =  0.01 $ and $\left| k_{\omega,Ge} -  \overline{k}_{\omega,Ge} \right| =\left| 0.01k_{\omega,Ge} - 0.009 \right|$. In addition, $\left| n_{\omega,GaAs} -  \overline{n}_{\omega,GaAs} \right| =  0.0065 $ and $\left| k_{\omega,GaAs} -  \overline{k}_{\omega,GaAs} \right| =\left| 0.03k_{\omega,GaAs} - 0.047 \right|$. Furthermore, $\left| n_{\omega,InP} -  \overline{n}_{\omega,InP} \right| =  \left| 0.01n_{\omega,InP} - 0.017 \right| $ and $\left| k_{\omega,InP} -  \overline{k}_{\omega,InP} \right| =\left| 0.01k_{\omega,InP} - 0.007 \right|$. Besides, in Table III, more accurate values of the regression coefficients of the neural networks are presented. Also, Table IV shows the values of $\theta_{r,x}$ and $\theta_{i,x}$ for semiconductors, i.e., Ge, GaAs, and InP. 

Based on the obtained results, the regression coefficients for the real part and the imaginary part are close to one, which shows the accuracy of the transfer learning-based strategy. To have a clearer understanding of the accuracy of the neural networks, Table V includes the quantitative validation performance of the ANNs. Here, the validation performance is formulated as the mean squared error of the validation dataset that is used during the training of the ANNs.

\section{Comparison}
\label{Sec:Discussion}

\begin{figure}
	\centering
	\includegraphics[width=1.05\linewidth] {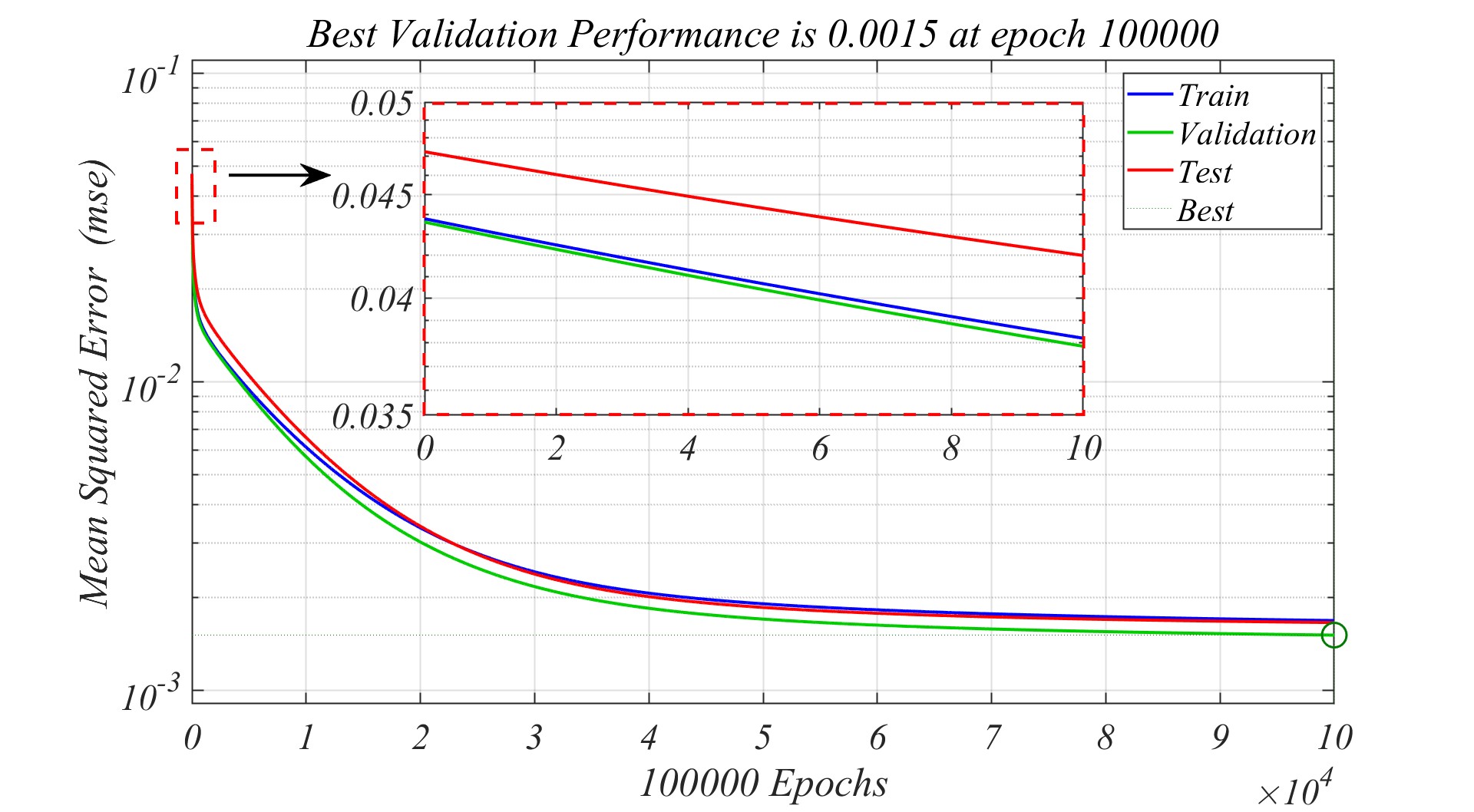}
	\caption{The performance plot related to the training of ANN\textsubscript{r,Au} with transfer learning strategy (the best validation performance is 0.0015).}
	\label{fig:v1}
\end{figure}

\begin{figure}
	\centering
	\includegraphics[width=1.05\linewidth] {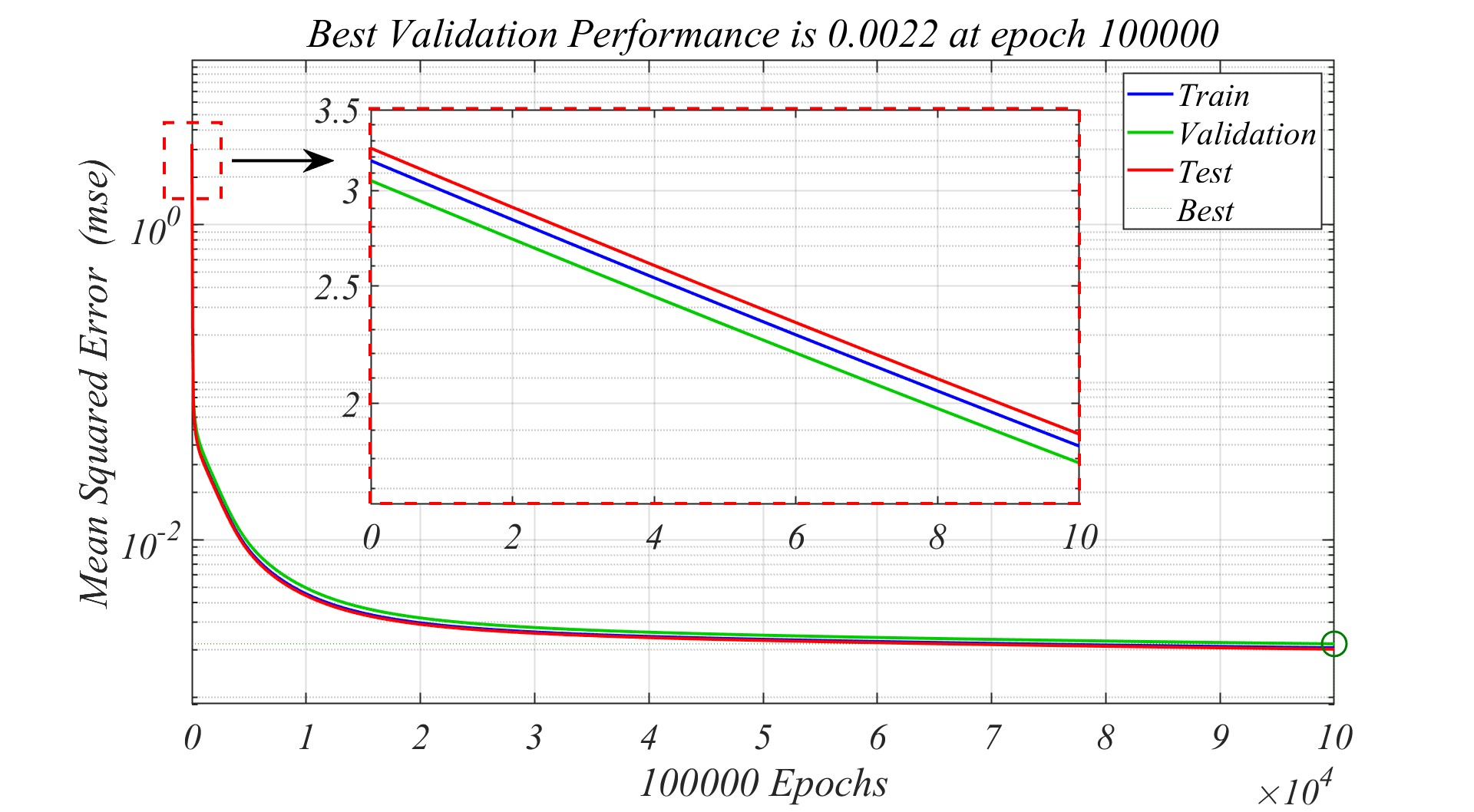}
	\caption{The performance plot related to the training of ANN\textsubscript{r,Au} without transfer learning strategy (the best validation performance is 0.0022).}
	\label{fig:v2}
\end{figure}

In this part, ANNs are trained without transfer learning and the results are compared with ANN\textsubscript{r,Au} and ANN\textsubscript{i,Au}, which has been trained using transfer learning: the results of the trained ANNs for the noble metals without the implementation of transfer learning are compared with the ANNs, that have been trained in Section III. To have a better understanding of the potential of transfer learning, the performances, i.e. cost function defined as the mean squared error (mse), of the ANNs will be discussed. In Fig. \ref{fig:v1} and Fig. \ref{fig:v2}, the performance of the ANNs to estimate the real part of the refractive index is shown. The corresponding ANN to Fig. \ref{fig:v1} is trained based on transfer learning. On the other hand, the corresponding ANN in Fig. \ref{fig:v2} is trained without the implementation of transfer learning.

Based on Fig. \ref{fig:v1}, the best validation performance under the transfer learning occurs at epoch 100000 with the value of 0.0015. Also, based on Fig. \ref{fig:v2}, the validation performance for the ANN without the deployment of transfer learning reached a value of 0.0022 at epoch 100000. In addition, based on Fig. \ref{fig:v1} and Fig. \ref{fig:v2}, the start point (i.e., epoch 0) for the validations of the ANNs with and without transfer learning are approximately 0.0436 and 3.0578, respectively. In other words, the ANN with transfer learning can start the training in a more accurate state (i.e., epoch 0), which can allow the training process to explore epochs that belong to more optimized values related to the structure of the ANN. In addition, based on Fig. \ref{fig:v1} and Fig. \ref{fig:v2}, the ANN with transfer learning can reach the validation performance with the value of 0.0022 at epoch 29719, and in other words, faster than the trained ANN without transfer learning.

\begin{figure}
	\centering
	\includegraphics[width=1.05\linewidth] {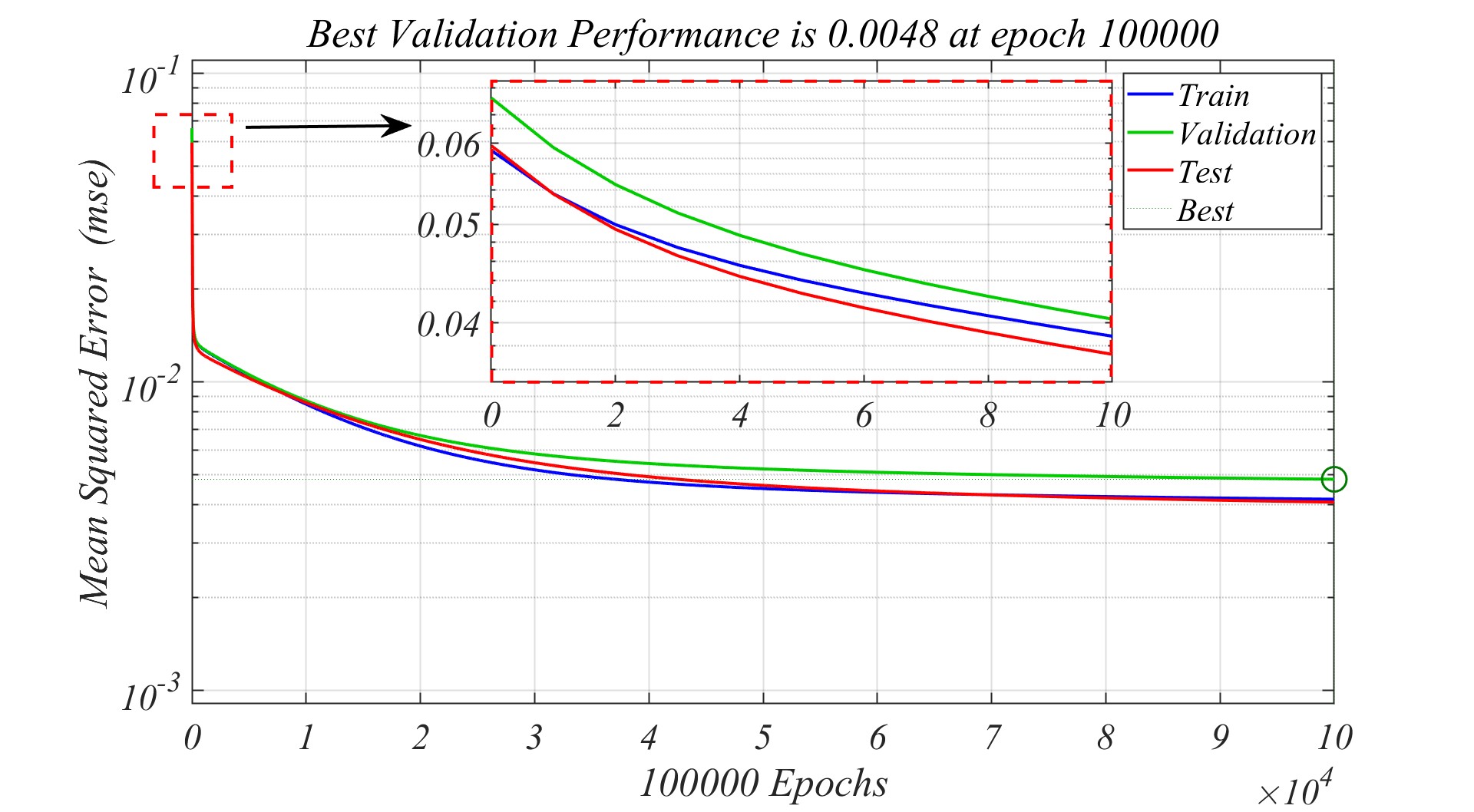}
	\caption{The performance plot related to the training of ANN\textsubscript{i,Au} with transfer learning strategy (the best validation performance is
0.0048).}
	\label{fig:v3}
\end{figure}

\begin{figure}
	\centering
	\includegraphics[width=1.05\linewidth] {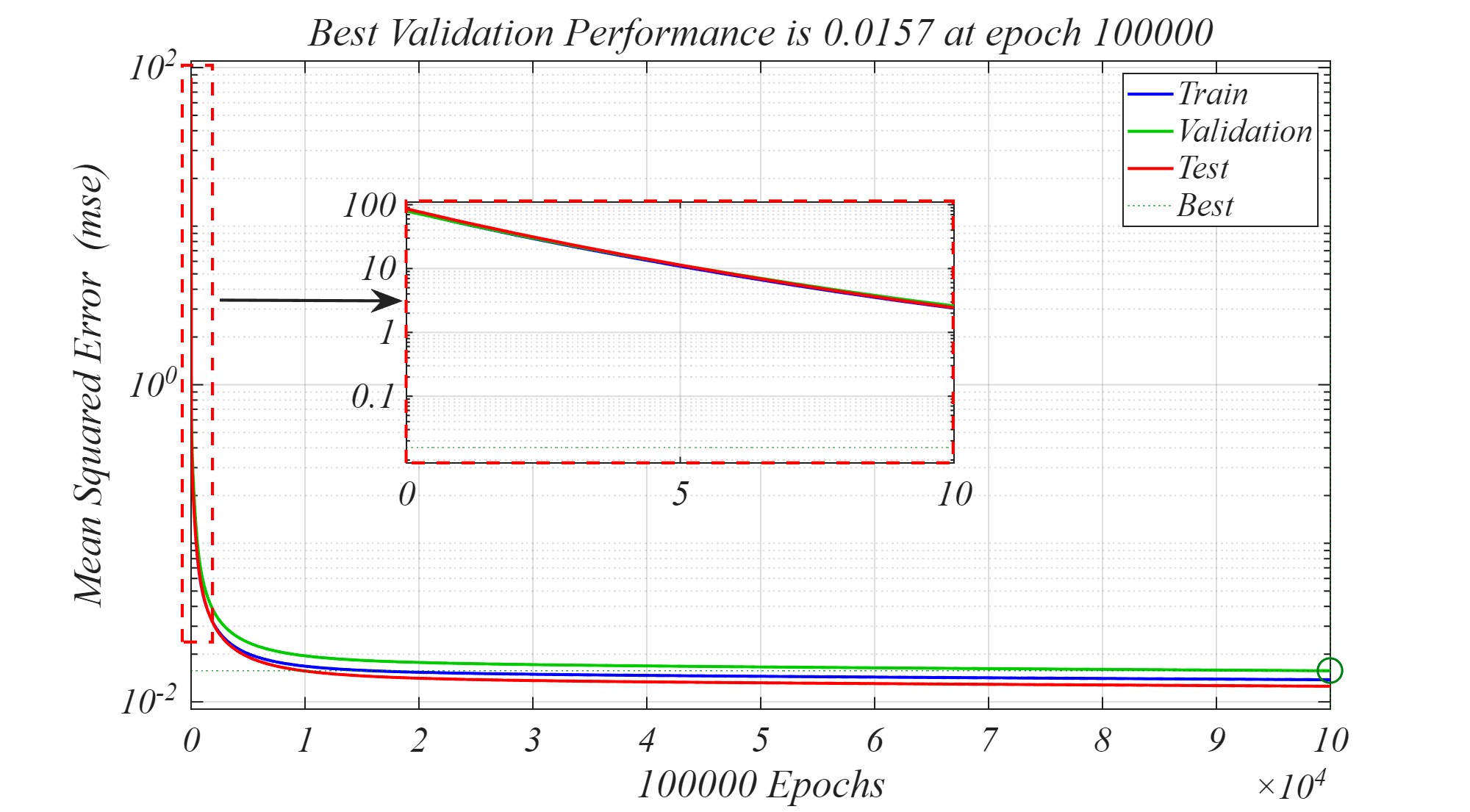}
	\caption{The performance plot related to the training of ANN\textsubscript{i,Au} without transfer learning strategy (the best validation performance is
0.0157).}
	\label{fig:v4}
\end{figure}

Furthermore, Fig. \ref{fig:v3} shows the performance the ANN trained based on transfer learning to calculate the imaginary value of the refractive index of gold. Based on Fig. \ref{fig:v3}, the best performance is related to epoch 100000 with a value of 0.0048. Also, the initial value of the performance of the ANN with transfer learning to estimate the imaginary part of the refractive index is 0.066 at epoch 0. Fig. \ref{fig:v4} illustrates the performance of the ANN that is trained without using transfer learning to calculate the value of the imaginary part of the refractive index. Based on Fig. \ref{fig:v4}, the training starts at epoch 0 with a performance of approximately 100. for this case study, the best validation happened at epoch 100000 with a value of 0.0157.

\begin{figure}
	\centering
	\includegraphics[width=1.05\linewidth] {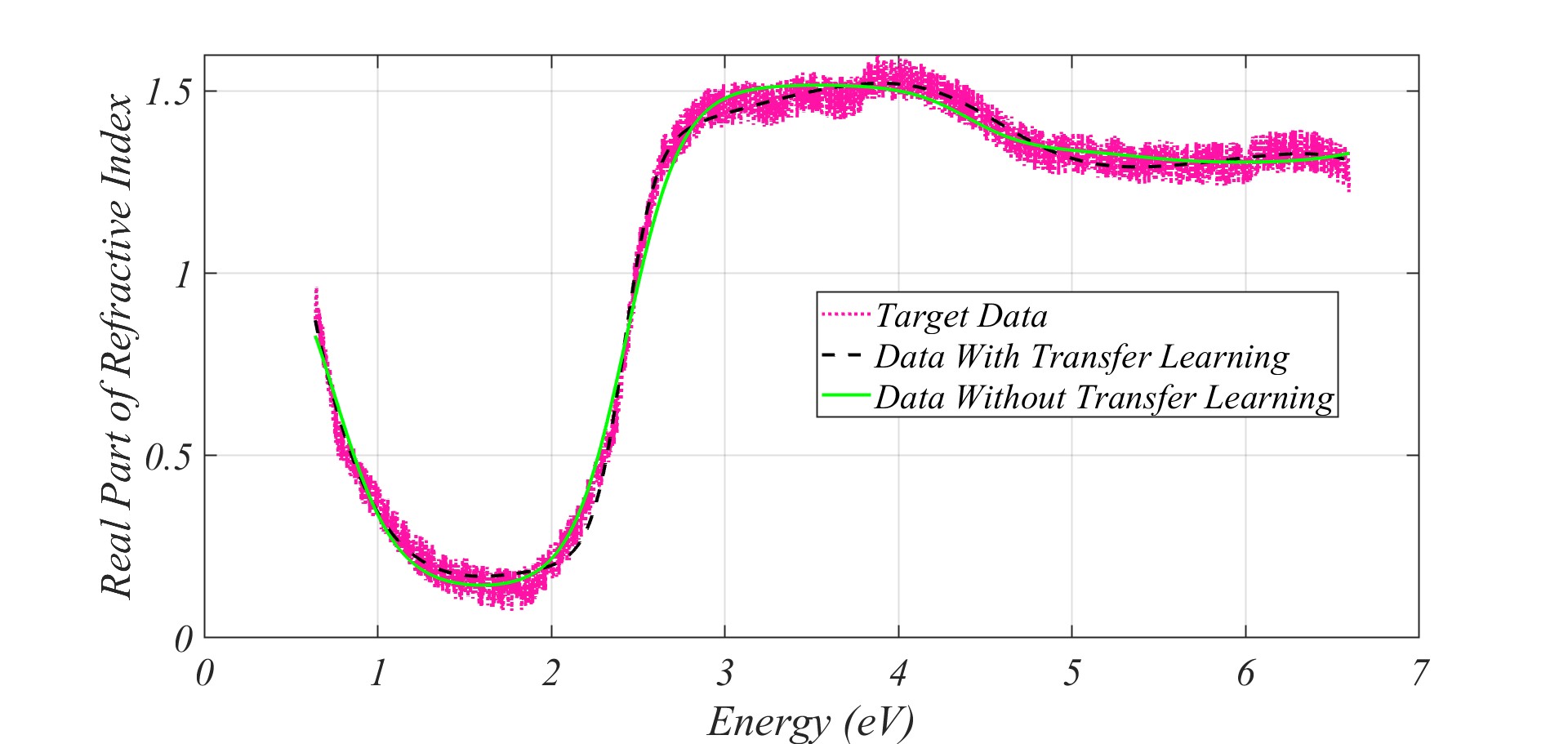}
	\caption{The comparison between neural networks (i.e., neural network with transfer learning and neural network without transfer learning) to estimate the real part of the refractive index of gold plotted as function of energy of light.}
	\label{fig:p1}
\end{figure}

\begin{figure}
	\centering
	\includegraphics[width=0.9\linewidth] {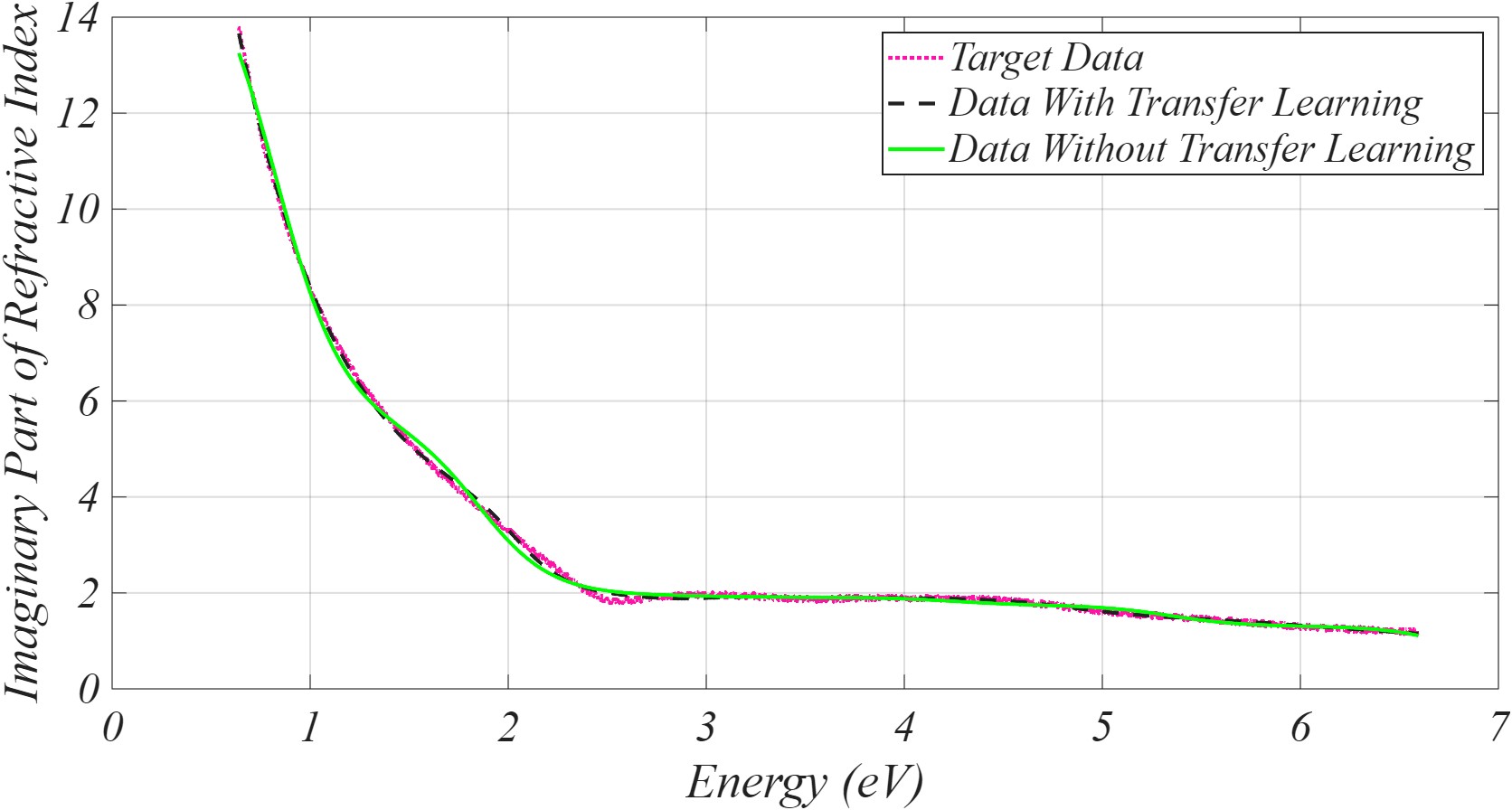}
	\caption{The comparison between neural networks (i.e., neural network with transfer learning and neural network without transfer learning) to estimate the imaginary part of the refractive index of gold plotted as function of energy of light.}
	\label{fig:p2}
\end{figure}

Also, Fig. \ref{fig:p1} and Fig. \ref{fig:p2} show the outputs of the ANNs and the target values. In particular, Fig. \ref{fig:p1} illustrates the value of the target (the real part of the refractive index), the estimated value using the ANN that is trained based on transfer learning, and the output of the ANN which is trained without the implementation of transfer learning. Fig. \ref{fig:p2} illustrates the target value of the imaginary part of the refractive index of gold, the output of the ANN that is trained based on transfer learning, and the ANN that is trained without the implementation of transfer learning. Based on Fig. \ref{fig:p1}, both ANNs can estimate the real part of the refractive index of gold, but, as mentioned above, the implementation of the transfer learning supports the training process to reach higher accuracy and in other words, fewer numbers of epochs (i.e., a faster process) while considering the same value of the performances. Besides, based on Fig. \ref{fig:p2}, both ANNs can successfully estimate the value of the target, but, the same as the real part, transfer learning for the imaginary part also leads to higher accuracy.

In summary, based on the results, the employment of transfer learning can support an ANN to have a more accurate trained neural network and, as a result, less error in the calculation. Additionally, for the training process, at a given epoch, the training of an ANN with the transfer learning can have a better performance compared to the training of the ANN without transfer learning. In other words, when defining a desired value for the performance of an ANN for the training, transfer learning can support the ANN trained with lower number of epochs (a faster training) compared to the case without the usage of the transfer learning. To have clearer comparison, Table. VI shows the parameters of the mentioned ANNs under transfer learning and without transfer learning to estimate the real part and the imaginary part of the refractive index of gold.

\begin{table*}[ht]
\centering
 \caption{The parameters of ANN\textsubscript{r,Au} and ANN\textsubscript{i,Au}, with transfer learning ($\checkmark$), and without transfer learning (X).}

 \begin{tabular}{c |c c | c c} 
 \hline
 \hline
  Transfer Learning & $\checkmark$  & $\checkmark$ & X & X \\  
 \hline
  Neural Network& ANN\textsubscript{r,Au}  & ANN\textsubscript{i,Au} & ANN\textsubscript{r,Au} & ANN\textsubscript{i,Au} \\  
 \hline
 Training mse & 0.0017 & 0.0042 & 0.0021 & 0.0138\\ 
 Validation mse & 0.0015 & 0.0048 & 0.0022 & 0.0157\\
  Test mse & 0.0017 & 0.0041 & 0.0020 & 0.0125\\
 Regression coefficient & 0.997 & 1 & 0.996 & 0.999\\
 Learning Rate & 0.01 & 0.001 & 0.01 & 0.001\\ 
Interpreter Line & $\overline{n}_{\omega,Au}=0.99 n_{\omega,Au}+0.0075$ & $\overline{k}_{\omega,Au}=k_{\omega,Au}+0.001$ & $\overline{n}_{\omega,Au}=0.99 n_{\omega,Au}+0.011$ & $\overline{k}_{\omega,Au}=k_{\omega,Au}+0.004$\\ 
 \hline
 \hline
 \end{tabular}
\end{table*}

\section{Study of the Signal-to-Noise Ratio}
\label{Sec:Noise}

In this part, a study on the noise under the implementation of the AI-based strategy is discussed. The value of the signal-to-noise ratio (SNR) is deployed as an index for the evaluation, and the study is done on Au. The value of $SNR$ for a signal can be defined as the ratio of the power of the signal to the power of the noise as follows \cite{matlabsnrnoise, matlabsnrnoise1}:

 \begin{equation}
SNR_d = 20 \times \log_{10}R_d,
\label{eqsnrrr}
 \end{equation} 
 and,
 \begin{equation}
R_d = \frac{\sqrt{\sum_{j=1}^{S_d} d(j)^2}}{\sqrt{\sum_{j=1}^{S_d} \tau(j)^2}}.
\label{eqsnrrr1}
 \end{equation} 

In \eqref{eqsnrrr}, $SNR_d$ is the value of $SNR$ for dataset $d$. In addition, in \eqref{eqsnrrr1}, $d(j)$ is the domain of the $j^{th}$ data, $\tau(j)$ is the noise in $j^{th}$ data, and $S_d$ is the size of the dataset. In this study,

 \begin{equation}
E_{r,x} =  \lvert n_{\omega_j,x} - CLN(n_{\omega_j,x}) \rvert = \lvert \tau_{\omega_j,r,x} \rvert,
\label{noise1}
 \end{equation} 

 and

 \begin{equation}
E_{r,ANN,x} =  \lvert \overline{n}_{\omega_j,x} - CLN(n_{\omega_j,x}) \rvert.
\label{noise2}
 \end{equation} 
 
In \eqref{noise1}, $\tau_{\omega_j,r,a}$ is the error or noise in the measured data (i.e., the real part of the refractive index) related to material $x$. Furthermore, $CLN(n_{\omega_j,x})$ is the real part of the refractive index of the material $x$ without any error or noise. Also, $\overline{n}_{\omega_j,x}$ is the output of $ANN_{r,x}$ to estimate the real part of the refractive index of the material $x$. Therefore, based on \eqref{noise1} and \eqref{noise2}, $E_{r,x}$ is corresponded to the domain of the error or noise of the dataset without using the AI-based technique. In addition, $E_{r,ANN,x}$ shows the error between the output of the ANN and the clean value of the dataset (i.e., without any error or noise). So, \eqref{noise2} shows the domain of the deviation of the estimated data from the original and clean data. In addition, 

 \begin{equation}
E_{i,x} =  \lvert k_{\omega_j,x} - CLN(k_{\omega_j,x}) \rvert = \lvert \tau_{\omega_j,i,x} \rvert,
\label{noise3}
 \end{equation} 

 and

 \begin{equation}
E_{i,ANN,x} =  \lvert \overline{k}_{\omega_j,x} - CLN(k_{\omega_j,x}) \rvert.
\label{noise4}
 \end{equation} 

In \eqref{noise3}, $\tau_{\omega_j,i,x}$ and $CLN(k_{\omega_j,x})$ are the error/noise in the dataset and the clean data, respectively and relate to the imaginary part of the refractive index of the material $x$. Besides, $\overline{k}_{\omega_j,x}$ is the output of the $ANN_{i,x}$ to calculate the imaginary part of the refractive index of the material $x$. Here, $E_{i,x}$ shows the absolute value of the error or noise of the dataset related to the imaginary part, and $E_{i,ANN,x}$ is is the absolute value of the deviation between the output of the ANN to calculate the imaginary part and the clean/original value of the mentioned data. 

\begin{figure}
	\centering
	\includegraphics[width=0.94\linewidth] {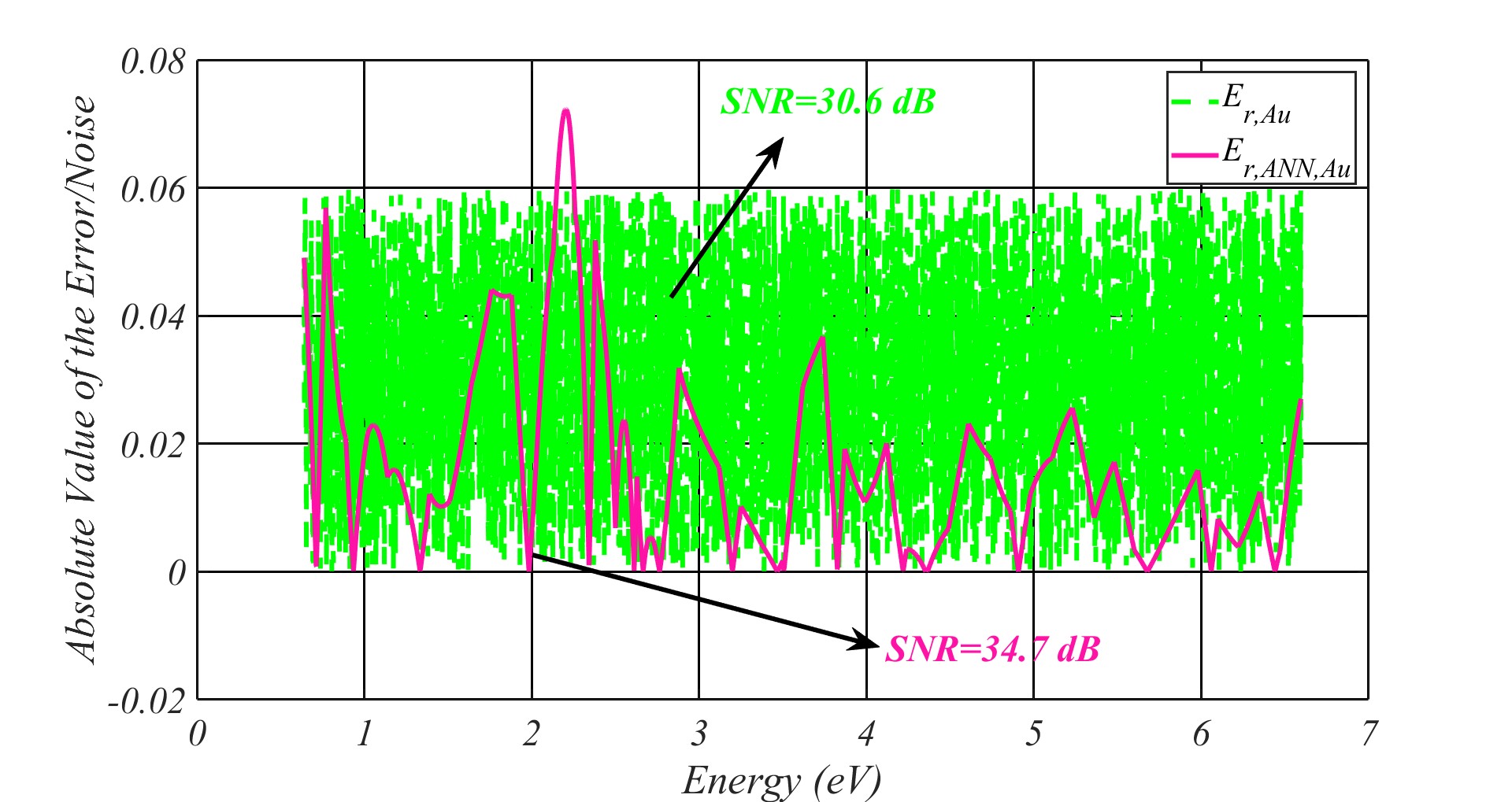}
	\caption{The absolute value of the error/noise related to the real part of the refractive index of Au plotted as function of energy of light. $E_{r,Au}$ is the absolute value of the implemented noise into the dataset related to Au (i.e., $E_{r,Au} =  \lvert n_{\omega_j,Au} - CLN(n_{\omega_j,Au}) \rvert = \lvert \tau_{\omega_j,r,Au} \rvert$), and $E_{r,ANN,Au}$ is the absolute value of the error between the output of the ANN and the clean/original real part of the refractive index without the implemented noise (i.e., $E_{r,ANN,Au} =  \lvert \overline{n}_{\omega_j,Au} - CLN(n_{\omega_j,Au}) \rvert$).}
	\label{fig:SNRAuR}
\end{figure}

\begin{figure}
	\centering
	\includegraphics[width=0.94\linewidth] {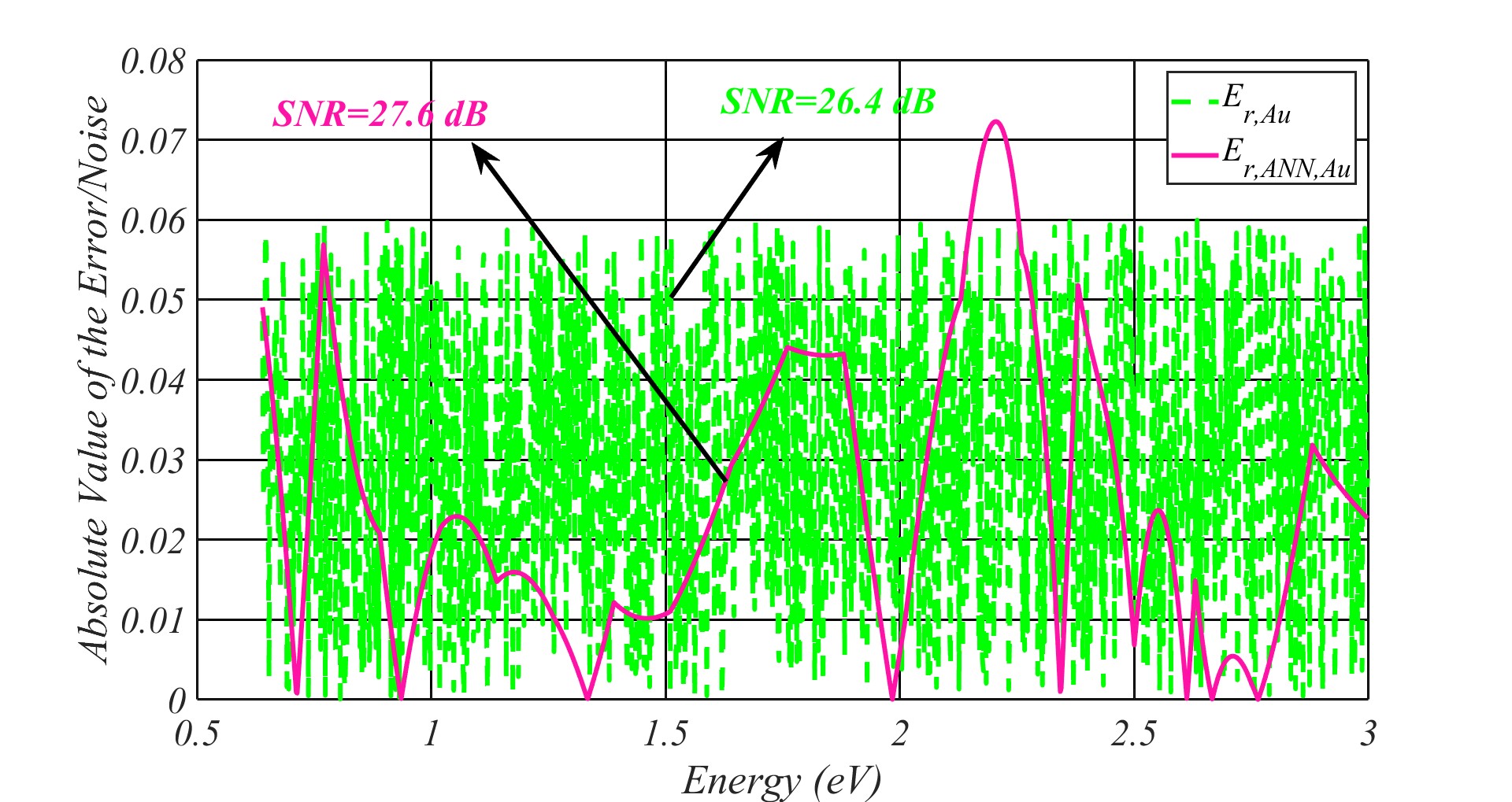}
	\caption{The absolute value of the error/noise related to the real part of the refractive index of Au plotted as function of energy of light for lower energies, i.e., less than $3eV$.}
	\label{fig:SNRAuR1}
\end{figure}

\begin{figure}
	\centering
	\includegraphics[width=0.94\linewidth] {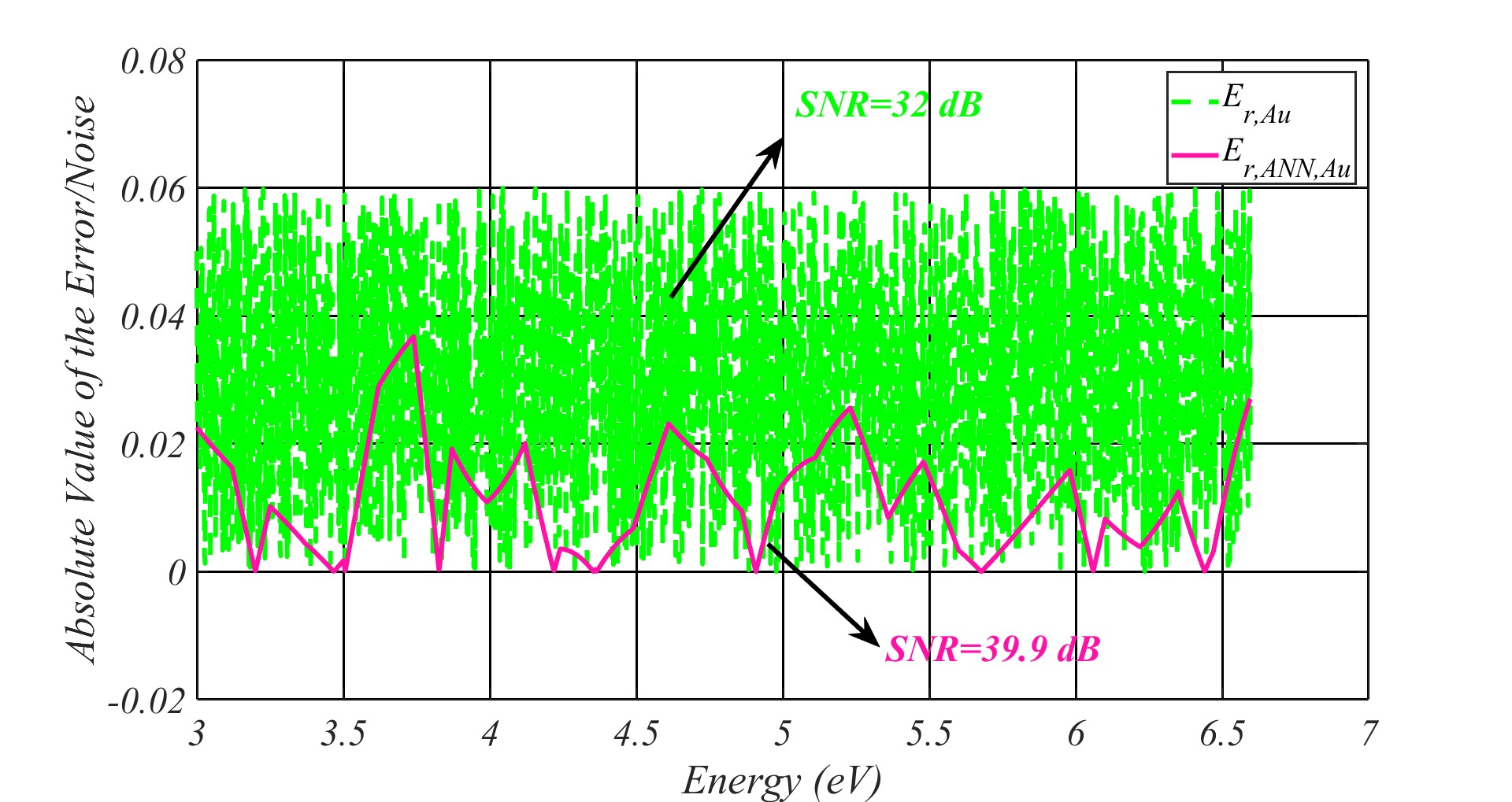}
	\caption{The absolute value of the error/noise related to the real part of the refractive index of Au plotted as function of energy of light for higher energies.}
	\label{fig:SNRAuR2}
\end{figure}

In Fig. \ref{fig:SNRAuR}, the absolute value of the error or noise of the real part of the refractive index of Au is depicted. Two parameters are deployed in Fig. \ref{fig:SNRAuR} to determine the absolute value of the error/noise, i.e., $E_{r,Au}$ and $E_{r,ANN,Au}$. Based on Fig. \ref{fig:SNRAuR}, the value of the $SNR$ related to the output of the ANN (i.e., $ANN_{r,Au}$) is 34.7, and the value of the $SNR$ corresponded to the noisy dataset is 30.6. Therefore, the value of the $SNR$ for the output of the ANN is more than the value of the $SNR$ of the noisy dataset that was implemented to train the ANN. So, considering that larger values of $SNR$ show a less-noisy dataset, it can be concluded that the output of the $ANN_{r,Au}$ is less-noisy than the used dataset. In other words, the implemented AI-based strategy could reduce the noise, i.e. it can make the dataset of the real part of the refractive index of Au cleaner. Furthermore, Fig. \ref{fig:SNRAuR1} and Fig. \ref{fig:SNRAuR2} illustrate the absolute value of the error/noise of the real part for the lower energies and the higher energies, respectively. Based on Fig. \ref{fig:SNRAuR1}, the value of $SNR$ related to the outputs of $ANN_{r,Au}$ and the implemented noisy dataset for lower energies (i.e., less than $3eV$) are 27.6 and 26.4, respectively. Besides, based on Fig. \ref{fig:SNRAuR2}, for the higher energies (i.e., more than $3eV$), the value of $SNR$ for the outputs of $ANN_{r,Au}$ and the noisy dataset are around 39.9 and 32, respectively. For both cases of lower and higher energies, the value of $SNRs$ of $ANN_{r,Au}$ are larger than the used noisy dataset, i.e., $ANN_{r,Au}$ could reduce the noise and make the real part of the refractive index of Au cleaner.


\begin{figure}
	\centering
	\includegraphics[width=0.94\linewidth] {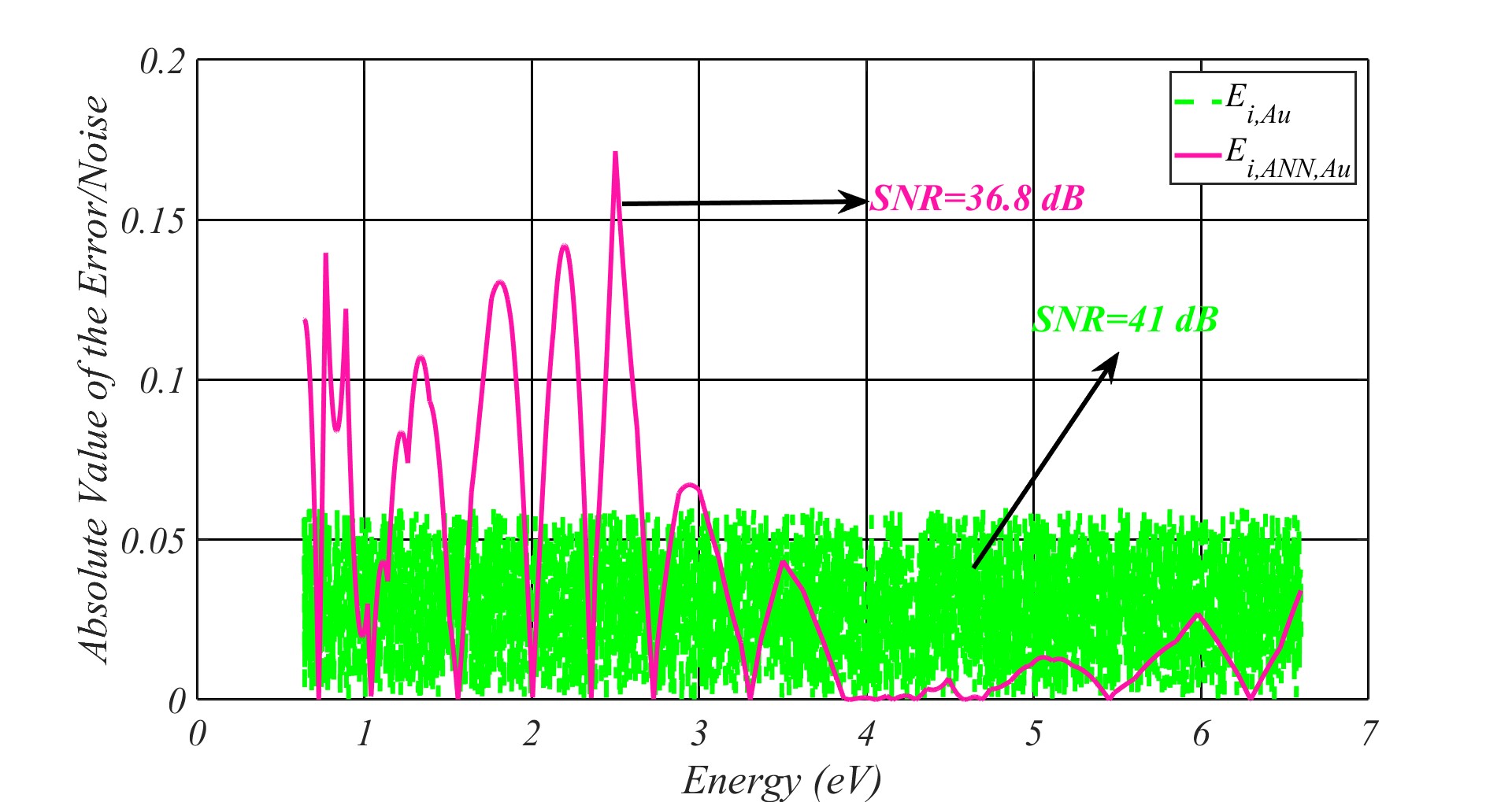}
	\caption{The absolute value of the error/noise related to the imaginary part of the refractive index of Au plotted as function of energy of light. $E_{i,Au}$ is the absolute value of the implemented noise into the dataset related to Au (i.e., $E_{i,Au} =  \lvert k_{\omega_j,Au} - CLN(k_{\omega_j,Au}) \rvert = \lvert \tau_{\omega_j,i,Au} \rvert$), and $E_{i,ANN,Au}$ is the absolute value of the error between the output of the ANN and the clean/original imaginary parr of the refractive index without the implemented noise (i.e., $E_{i,ANN,Au} =  \lvert \overline{k}_{\omega_j,Au} - CLN(k_{\omega_j,Au}) \rvert$).}
	\label{fig:SNRAuI}
\end{figure}

\begin{figure}
	\centering
	\includegraphics[width=0.94\linewidth] {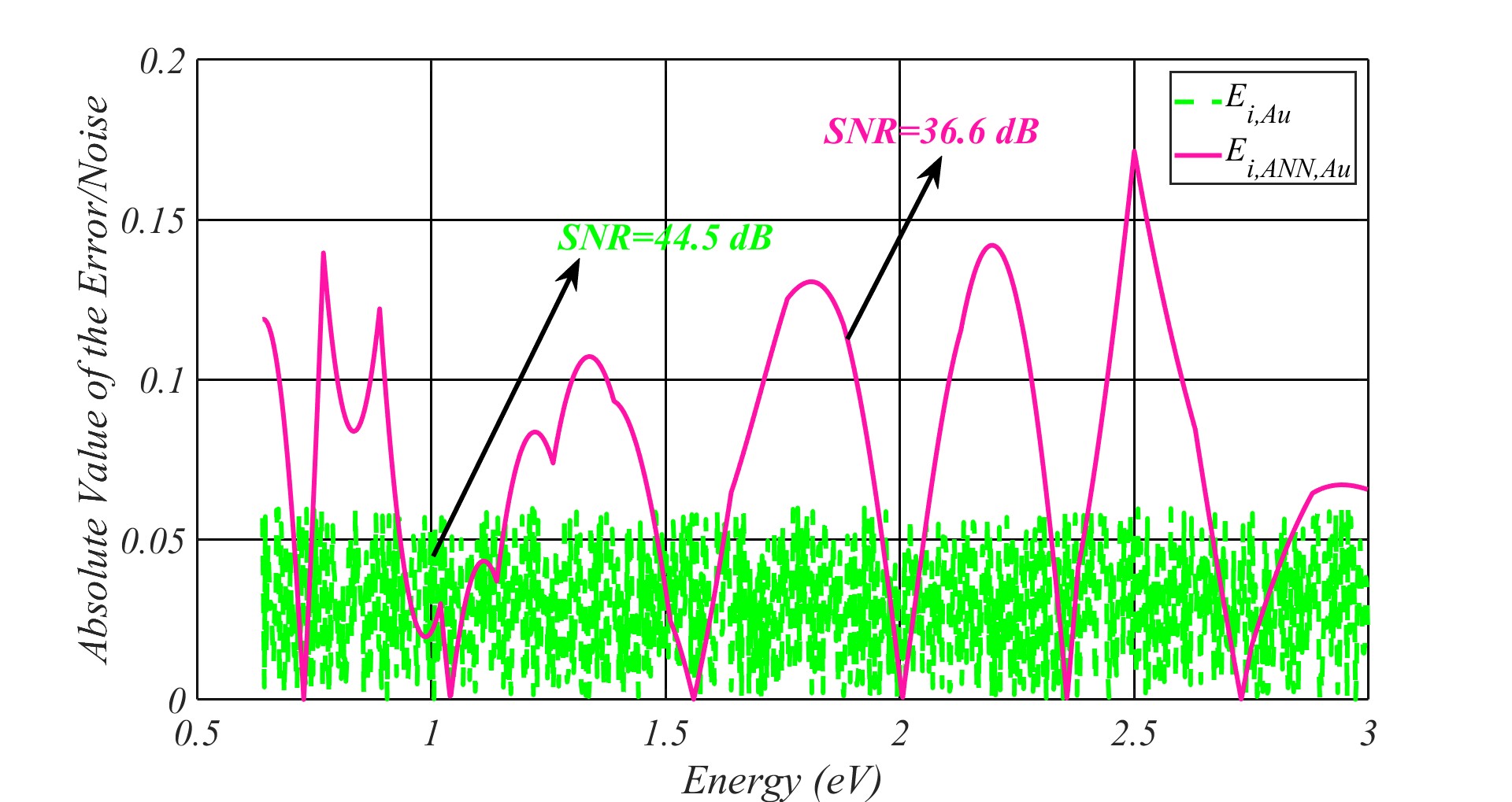}
	\caption{The absolute value of the error/noise related to the imaginary part of the refractive index of Au plotted as function of energy of light for lower energies, i.e., less than $3eV$.}
	\label{fig:SNRAuI1}
\end{figure}

\begin{figure}
	\centering
	\includegraphics[width=0.94\linewidth] {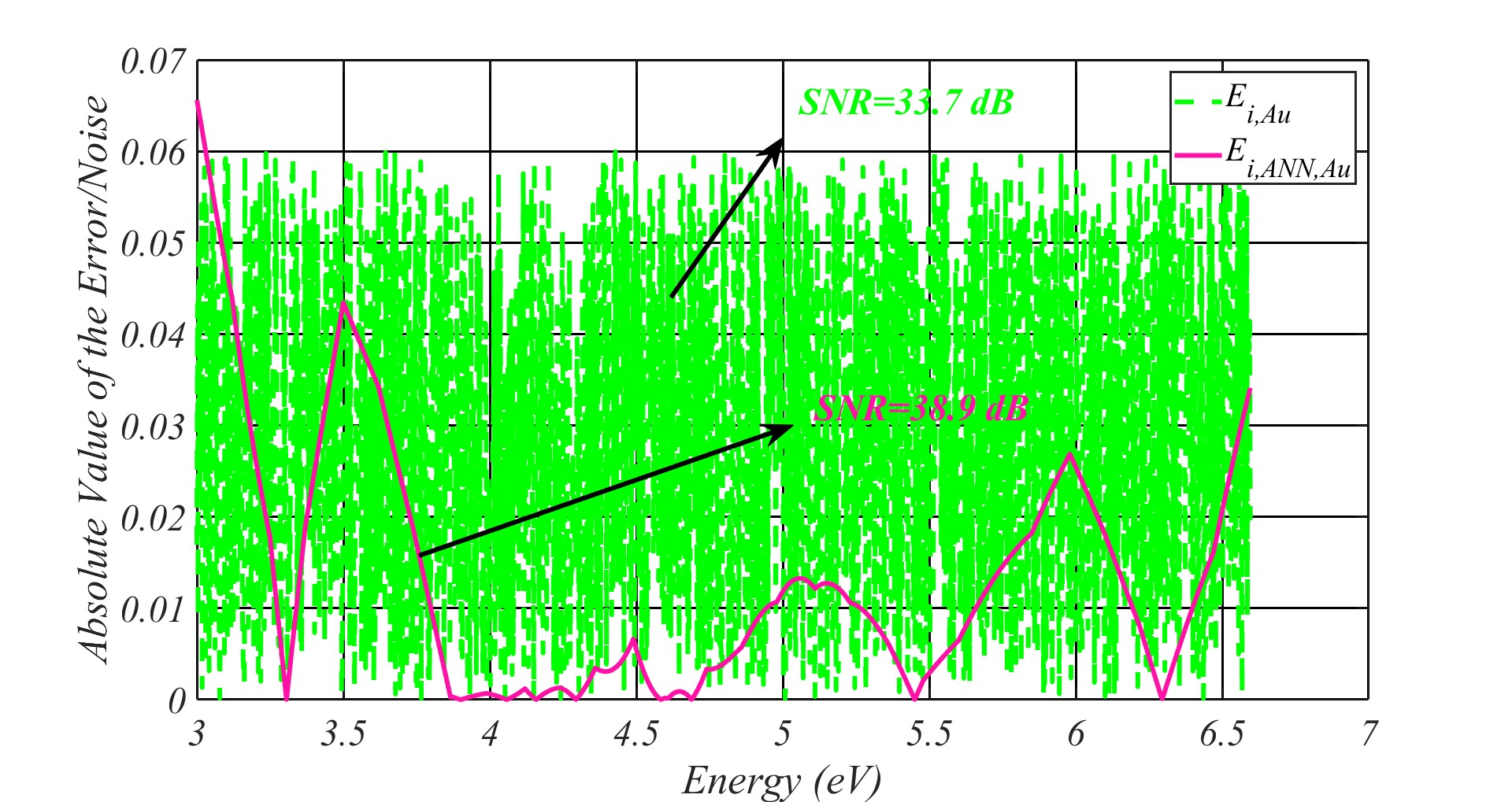}
	\caption{The absolute value of the error/noise related to the imaginary part of the refractive index of Au plotted as function of energy of light for higher energies.}
	\label{fig:SNRAuI2}
\end{figure}

Additionally, the absolute values of the error or noise corresponding to the imaginary part of the refractive index of Au are shown by Fig.\ref{fig:SNRAuI}, Fig.\ref{fig:SNRAuI1}, and Fig.\ref{fig:SNRAuI2}. Based on Fig. \ref{fig:SNRAuI}, the values of $SNR$ for the outputs of $ANN_{i,Au}$ and the noisy dataset are about 36.8 and 41, respectively. Furthermore, based on Fig. \ref{fig:SNRAuI1}, for the lower energies, the value of $SNR$ related to the outputs of $ANN_{i,Au}$ is smaller than the noisy dataset. On the other hand, based on Fig. \ref{fig:SNRAuI2}, for higher energies, the value of $SNR$ corresponded to $ANN_{i,Au}$ is larger than the $SNR$ of the noisy data. Therefore, based on Fig.\ref{fig:SNRAuI}, Fig.\ref{fig:SNRAuI1}, and Fig.\ref{fig:SNRAuI2}, for higher energies, the AI-based technique can reduce the noise of the imaginary part of the refractive index of Au for the noisy dataset.


\begin{figure}
	\centering
	\includegraphics[width=1.084\linewidth] {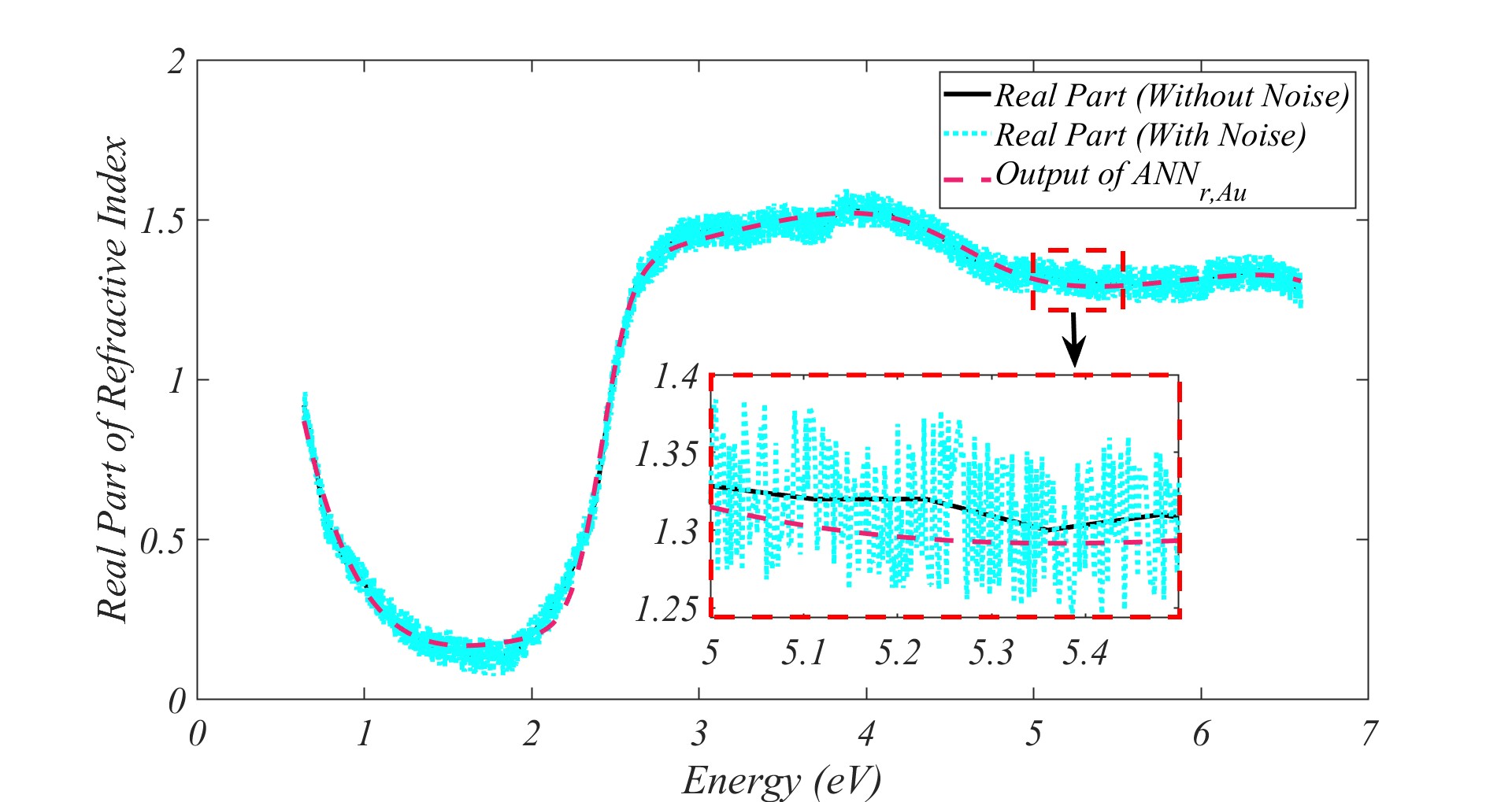}
	\caption{The real part of the refractive index of Au plotted as function of energy of light.}
	\label{fig:allfigsAuR}
\end{figure}

\begin{figure}
	\centering
	\includegraphics[width=1.084\linewidth] {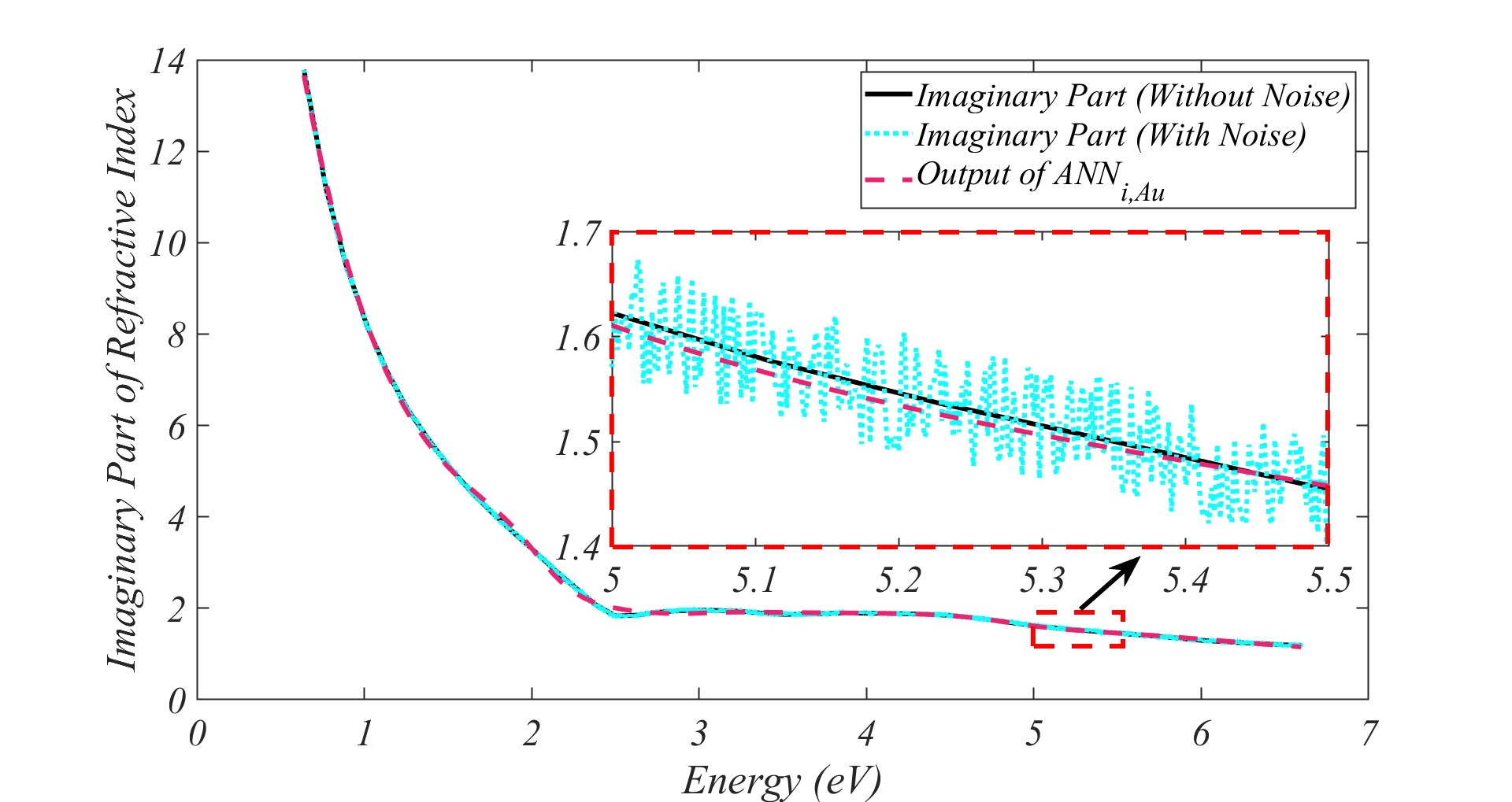}
	\caption{The imaginary part of the refractive index of Au plotted as function of energy of light.}
	\label{fig:allfigsAuI}
\end{figure}

To conclude, Fig. \ref{fig:allfigsAuR} shows the real part of the refractive index of Au including clean dataset without the error/noise, noisy dataset, and the estimated values by $ANN_{r,Au}$. As it is shown in Fig. \ref{fig:allfigsAuR}, the output of the ANN has less noise and as a result cleaner data than the noisy dataset. Based on Fig. \ref{fig:allfigsAuR}, the outputs of $ANN_{r,Au}$ have less fluctuations around the real data, and the noisy dataset has larger fluctuations around the real data caused by the noise. Furthermore, Fig. \ref{fig:allfigsAuI} depicts the imaginary part of the refractive index of Au containing dataset without the noise, noisy dataset, and the output of $ANN_{i,Au}$ used to calculate the imaginary part. Based on Fig. \ref{fig:allfigsAuI}, the noisy dataset has more fluctuations when it is compared to the output of $ANN_{i,Au}$. Finally, based on the obtained results, $ANN_{r,Au}$ can estimate the real part of the refractive index of Au with less noise compared to the noisy dataset. Also, for higher energies, $ANN_{i,Au}$ can estimate the imaginary part of the refractive index of Au with less noise comparing to the noisy dataset, i.e., it can produce cleaner data for higher energies. The values of $SNRs$ related to different ranges of energies for the real part and the imaginary part of the refractive index of Au are shown in Table VII and Table VIII, respectively.

\begin{table}[]
\centering
 \caption{The value of $SNR$ related to the noisy dataset and the outputs of $ANN_{r,Au}$ for the real part of the refractive index of Au.}
\begin{tabular}{c |c c c}
SNR & All energies & Low Energies  & High Energies \\ \hline
\hline
Noisy Dataset             & 30.6             & 26.4  & 32                      \\
Outputs of $ANN_{r,Au}$           & 34.7               & 27.6      & 39.9                   \\ \hline \hline
Noise reduction \\ by $ANN_{r,Au}$           & $\checkmark$               & $\checkmark$      & $\checkmark$
\end{tabular}
\end{table}

\begin{table}[]
\centering
 \caption{The value of $SNR$ related to the noisy dataset and the outputs of $ANN_{i,Au}$ for the imaginary part of the refractive index of Au.}
\begin{tabular}{c |c c c}
SNR & All energies & Low Energies & High Energies \\ \hline
\hline
Noisy Dataset             & 41             & 44.5  & 33.7                      \\
Outputs of $ANN_{i,Au}$           & 36.8             & 36.6      & 38.9                   \\     \hline \hline
Noise reduction \\ by $ANN_{i,Au}$           & $-$               & $-$      & $\checkmark$                  
\end{tabular}
\end{table}

\section{Ablation Study}
\label{Sec:Ablation Study}
To provide a clear understanding of the influence of transfer learning, an ablation study was conducted. Three case studies for a given material are considered: Random, RTR (relevant transfer learning), and N-RTR (non-relevant transfer learning). For the first case study, the neural networks for both the real and imaginary parts of the refractive index is trained with random initializations. For RTR, the initialization of the neural networks is made using parameters transferred from physically related materials. Finally, initialization using parameters transferred from an unrelated material for the neural networks is considered. We are considering Au as the given material, Cu as the physically related material, and Si as an unrelated material. In addition, for both the imaginary and the real parts of the refractive index of Au, random numbers between -0.1 and 0.1 are added to the original dataset to play the role of noise or the error of the measurements. Also, maximum numbers of epochs for all case studies are set to 10000, and the same architecture for all neural networks are deployed. 

\begin{figure}
	\centering
	{\includegraphics[width=1\linewidth] {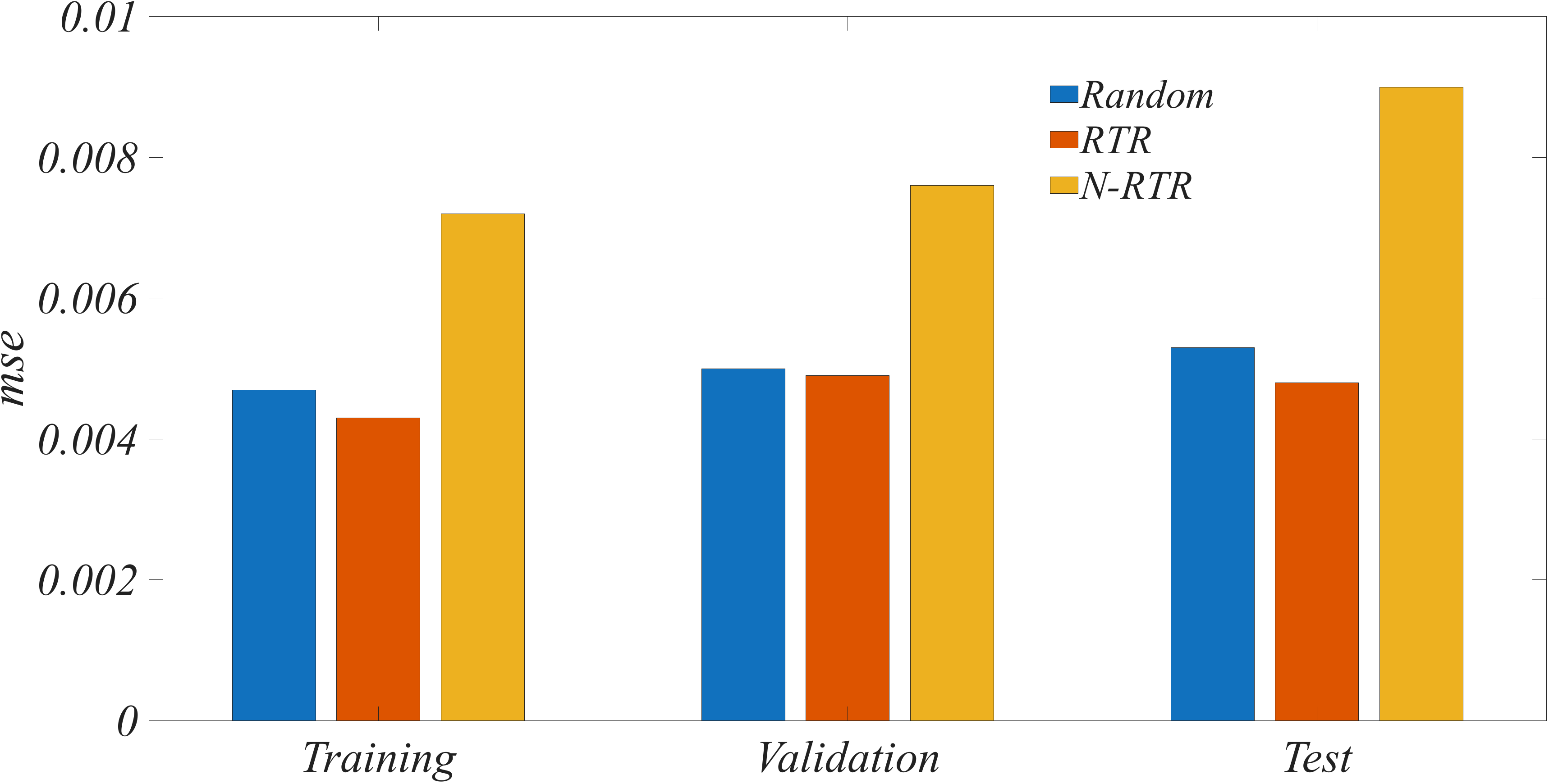}
    }
	\caption{The value of mse for the training, validation, and testing of the neural networks corresponding to ablation case studies and related to the real part of the refractive index of Au.}
	\label{fig:Ablation_real}
\end{figure}

\begin{figure}
	\centering
	{\includegraphics[width=1\linewidth] {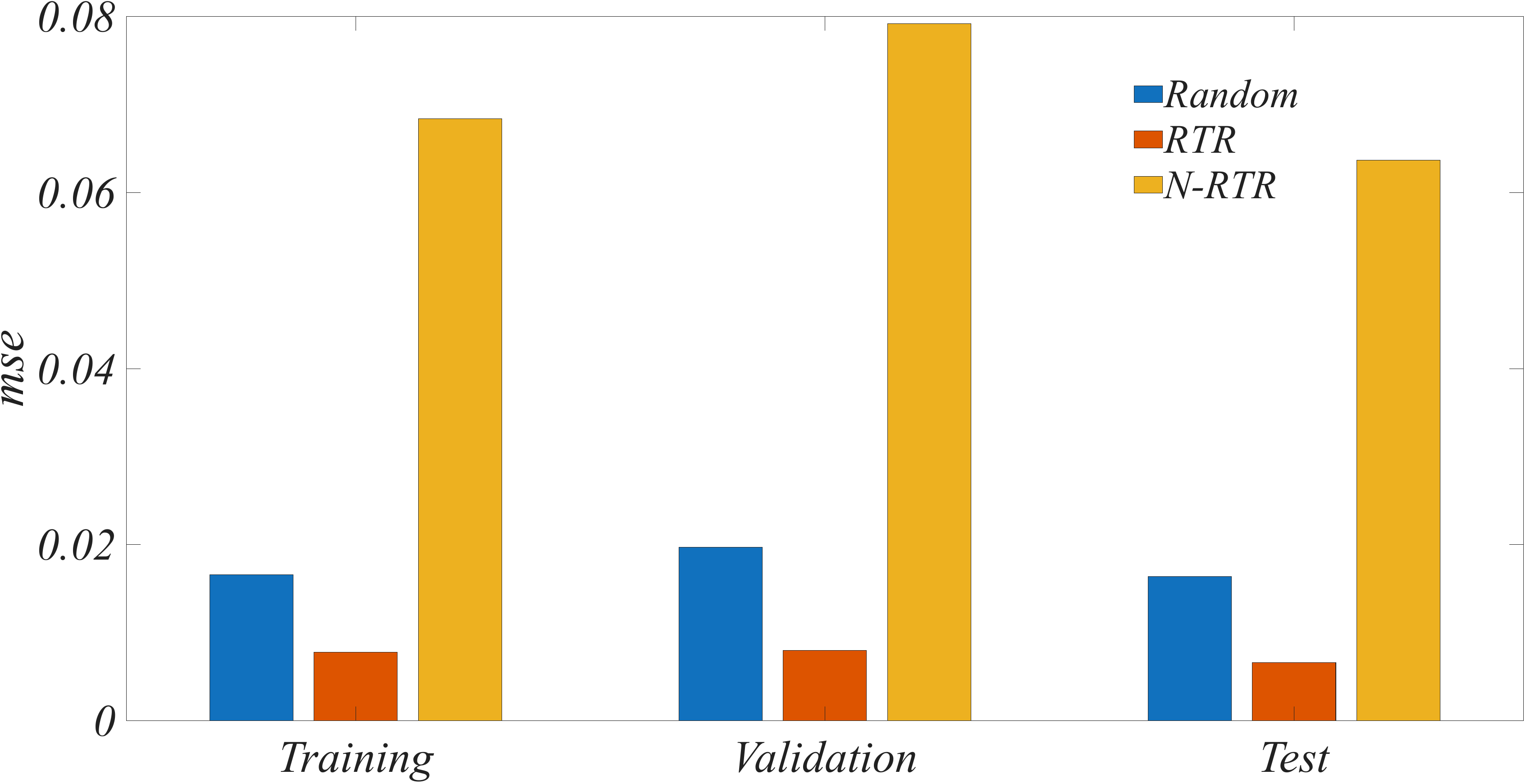}
    }
	\caption{The value of mse for the training, validation, and testing of the neural networks corresponding to ablation case studies and related to the imaginary part of the refractive index of Au.}
	\label{fig:Ablation_imaginary}
\end{figure}

The results that we obtain are shown in Fig. \ref{fig:Ablation_real} and Fig. \ref{fig:Ablation_imaginary}, for the real and imaginary part of the refractive index, respectively. Based on Fig. \ref{fig:Ablation_real} and Fig. \ref{fig:Ablation_imaginary}, the minimum mean squared error values are achieved under RTR, and the highest values under N-RTR: transferring knowledge from a physically related material could decrease the error of the calculations, while transferring the knowledge from the unrelated material would enlarge the calculation errors. The mean squared error values for the mentioned three case studies are shown explicitly in Table. IX and X.
The errors obtained are significantly larger than the ones obtained from our transfer-learning approach, thus proving the validity and effectiveness of the novel approach that we have developed.

\begin{table}[]
\centering
 \caption{Ablation Study consists of three case studies for the real part of the refractive index of Au with learning rate 0.03 for all three case studies corresponding to Fig. \ref{fig:Ablation_real}.}
\begin{tabular}{c |c c c }
\rotatebox{0}{Case study} & \rotatebox{0}{Training mse} & \rotatebox{0}{Validation mse} & \rotatebox{0}{Test mse}  \\ \hline
\hline
Random             & 0.0047             & 0.0050  & 0.0053                   \\ \hline
RTR           & 0.0043             & 0.0049      & 0.0048                 \\     \hline
N-RTR            & 0.0072               & 0.0076      & 0.0090                
\end{tabular}
\end{table}

\begin{table}[]
\centering
 \caption{Ablation Study consists of three case studies for the imaginary part of the refractive index of Au with learning rate 0.003 for all three case studies corresponding to Fig. \ref{fig:Ablation_imaginary}.}
\begin{tabular}{c |c c c }
\rotatebox{0}{Case study} & \rotatebox{0}{Training mse} & \rotatebox{0}{Validation mse} & \rotatebox{0}{Test mse}  \\ \hline
\hline
Random             & 0.0166             & 0.0197  & 0.0164                   \\ \hline
RTR           & 0.0078             & 0.0080      & 0.0066                 \\     \hline
N-RTR            & 0.0684               & 0.0792      & 0.0637                
\end{tabular}
\end{table}

\section{Discussion}
\label{Sec:Discussionii}

Over the last few years, few attempts have been made to use AI for the estimation of the refractive index of materials. Previous works used complex AI-based models (e.g., deep networks or a lot of calculation units such as neurons) or implemented more than one input for the model \cite{mcatee2019artificial, mcatee2019artificialP, esmaeili2023insights, palichev2023prediction, ghorbani2024general}. Deployment of a complex model can have disadvantages meaning that it can be a challenging task to be used in real applications. In addition, due to the number of the hidden layers and the inputs required, they can be computationally costly to achieve an accurate trained model. Furthermore, generally, complex deep learning-based models need a large dataset for the training. To overcome the mentioned challenges (i.e., complexity of the model, computational cost, and dependency to the size of the dataset to achieve a good level of the accuracy), in this study, transfer learning has been implemented to train ANNs (with only one input and one hidden layer) to estimate the real part and the imaginary part of the refractive index of different materials. For this purpose, two well-trained ANNs have been implemented to transfer their knowledge to other agents (i.e., materials) to initialize their parameters and train the ANNs based on the initialized values of the parameters. Two types of materials, i.e., noble metals, and semiconductor materials are considered. For the case of noble metals, the knowledge of the ANNs of Cu has been transferred to use for the training of the ANNs for Au. In addition, for the case of semiconductor materials, the knowledge of the ANNs for Si has been used to tune the parameters of the ANNs for Ge, GaAs, and InP. Uniform random numbers have been added to the original dataset to play the role of the measurement error or noises in the system to evaluate the performance of the transfer learning under a worse situation. For the case of noble metals, a data generation strategy has been deployed to increase the size of the dataset.

For noble metals, the implementation of transfer learning allowed for a more accurate refractive index estimation compared to ANNs without transfer learning. For example, the validation performances (i.e., mse) for the case using transfer learning were 0.0015 and 0.0048 for the real part and the imaginary part of the refractive index, respectively. On the other hand, the validation performances related to the real part and the imaginary part without the transfer learning were 0.0022 and 0.0157, respectively. In addition, for the case of semiconductor materials, only the original dataset with the added noises or errors has been implemented, which has a small size and only included 46 samples for each material. In this case, a dataset with a small size, transfer learning-based strategy provided a very good level of accuracy in calculating the value of the refractive index, including both the real part and the imaginary part. For the real part of the refractive index of Ge, GaAs, and InP, the validation performance were 0.0093, 0.0042, and 0.0036, respectively. Also, the validation performance of the calculation of the imaginary part of Ge, GaAs, and InP were 0.0167, 0.0376, and 0.0037, respectively. The noise reduction capability of the implemented strategy is examined using the SNR index. This evaluation has been done on dataset related to Au. The implemented AI-based method could improve the SNR value corresponding to the real part of the refractive index of Au. In addition, for the imaginary part of the refractive index of Au, the AI-based model could enhance that value for higher energies, i.e., more than $3eV$.

\section{Conclusions and Future Work}
\label{Sec:Future}

Many possible directions can be suggested to deploy AI for classical and quantum physics-based research lines, e.g., classical photonics and quantum optics. The refractive index of the materials is a critical parameter for the quantum photonic devices simulations, particularly for cavities with high quality factor. In the case of small discrepancies between the real and the implemented values of the refractive index, cavity modes may not be in resonance with single-photon emitters, thus negatively affecting device performances. Information about the refractive index is crucial for research related to optical resonators or cavity \cite{sapienza2008electrically, trojak2017metallic, Daniel:11, Tsigaridas, 11062684}, waveguides \cite{trojak2017aaa, davanco2017heterogeneous, KAZANSKIY2021106863, Cheben:10}, and single quantum dots \cite{sapienza2015nanoscale, CAKIR20122659}, key ingredients for quantum technology applications. The refractive index is a parameter essential to evaluate and control the interaction between light and matter, and the improvement of this interaction is vital for quantum and classical photonics \cite{trojak2021cavity}. 

This study, that investigates example of metals and semiconductor materials, shows a novel approach for the accurate evaluation of the refractive index using transfer learning: a data-driven curve fitting of the real and imaginary part of the refractive index of materials, with the learned ANN parameterization for one material transferred as an initialization for fitting another material. Its application can allow the development and investigation of novel materials, such as new compound materials, complex materials, thin films, etc., where partial data is existent and can represent the basis for the training of the AI. Furthermore, this model could be extended and applied to evaluate additional variables like wavelength, temperature dependency, composition, or other material descriptors specific to the devices under study.


\textbf{Acknowledgments:} LS acknowledges financial support from the Engineering and Physical Sciences Research Council, grant EP/X040526/1.

\textbf{Dataset:} The original datasets for noble metals and semiconductor materials can be found in \cite{PhysRevB.6.4370}, and \cite{PhysRevB.27.985}, respectively.



\bibliography{Reference}
\bibliographystyle{ieeetr}
\end{document}